\documentclass[12pt]{article}
\usepackage{latexsym, amsmath, epsfig, amsfonts, amsbsy}
\DeclareFontFamily{OT1}{rsfs10}{}
\DeclareFontShape{OT1}{rsfs10}{m}{n}{ <-> rsfs10 }{}
\DeclareMathAlphabet{\mathscript}{OT1}{rsfs10}{m}{n}

\else\oddsidemargin25mm\evensidemargin25mm\marginparwidth25mm\fi
\def\Z{\mathbb{Z}}
\def\C{\mathbb{C}}

\def\R{\mathbb{R}}

\def\bpl{\Big(}
\def\bpr{\Big)}

\newcommand{\ft}[2]{{\textstyle\frac{#1}{#2}}}
\def\brr{\begin{equation}}
\def\err{\end{equation}}
\def\brr{\begin{eqnarray}}
\def\err{\end{eqnarray}}
\def\ba{\left(\begin{array}}
\def\ea{\end{array}\right)}
\def\lf{\left.\begin{array}{c}}
\def\rf{\end{array}\right.}

\newcommand{\dr}{\raise.3ex\hbox{$\stackrel{\leftarrow}{\partial }$}{}}
\newcommand{\dl}{\raise.3ex\hbox{$\stackrel{\rightarrow}{\partial}$}{}}
\newcommand{\topi}{\raise.3ex\hbox{$\stackrel{\pi}{\longrightarrow}$}{}}
\newcommand{\ns}{\normalsize}

\renewcommand{\a}{\alpha}
\renewcommand{\b}{\beta}

\newcommand{\g}{\gamma}
\newcommand{\G}{\Gamma}

\newcommand{\q}{\theta}

\begin{document}


\begin{titlepage}

\vspace{-4cm}

\title{
   \hfill{\ns CU-TP-1064\,,HWS-200201\,,HU-EP-02/28\,\\}
   \hfill{\ns hep-th/0207162\\[2cm]}
   {\LARGE Intersecting Branes in {\it M}-Theory \\
   and Chiral Matter in Four Dimensions \\[1.5cm]  }}

\author{{\bf
   Charles F. Doran$^{1}$
   and
   Michael Faux$^{1,2}$}\\[5mm]
   {\it $^1$Departments of Mathematics and Physics} \\
   {\it Columbia University} \\
   {\it 2990 Broadway, New York, NY 10027} \\[3mm]
   {\it $^2$ Department of Physics} \\
   {\it Hobart and William Smith Colleges} \\
   {\it Geneva, NY 14456}}
\date{}

\maketitle

\vspace{.3in}

 \begin{abstract}
 \noindent
 We explicitly derive a complementary pair of four-dimensional
 {\it M}-theory brane-world models, linked by a five-dimensional
 bulk, each of which has a unique anomaly-free chiral spectrum.
 This is done via resolution of local consistency
 requirements, in the context of the simplest
 global quotient $T^7/\Gamma$ involving ten-dimensional fixed-planes,
 for which a chiral four-dimensional spectrum could arise.

 \vspace{.5in} \noindent
 \end{abstract}

\thispagestyle{empty}

\end{titlepage}


 \section{Introduction}
 One of the preeminent tasks of contemporary theoretical physics is to
 seek a mathematically consistent higher-dimensional explanation
 for the chiral fermion spectrum and gauge symmetries of the
 standard model.  Over the last decade,
 string theory has precipitated a virtual miasma
 of related ideas.  Recently, two different sorts of constructions
 have emerged as compelling avenues for the derivation of effective
 physics from within both string theory and also its elusive non-perturbative
 cousin, {\it M}-theory.  On the one hand, brane-world models
 \cite{bkl1, bkl2, csu, afiru1}, obtained
 by consistent inclusion of intersecting D-branes and open strings
 in various background geometries
 \cite{afiru2, b2kl}, have succeeded in providing
 a plausible context for the standard model itself
 \cite{imr, bwstd, csu1, csu2}.  On the other
 hand, {\it M}-theory has inspired a search for a more
 elegant eleven-dimensional underpinning to some of these
 same constructions, and has stimulated an interest in the
 physically-relevant mathematical
 characterization of $G_2$ holonomy seven-manifolds \cite{Ach,AtWit}.

 There are two essential obstructions which have hampered the
 search for {\it M}-theoretic phenomenology
 as compared to string theory analogues.  One is the fact
 that a fundamental description of {\it M}-theory has not yet emerged.
 The other is that much less is known about $G_2$ holonomy seven-manifolds
 than is known about Calabi-Yau threefolds.  Thus, it is
 difficult to provide geometric explanations for the symmetries
 and spectra which may arise in {\it M}-theory compactifications.
 However, there is one restricted class of constructions
 tailor-made to shed light on each of these two problems,
 which also includes a built-in mechanism for resolving effective physics.  This is the
 class of models based on global orbifold compactifications of
 eleven-dimensional supergravity.

 One reason why orbifold compactifications are so useful in {\it M}-theory
 is that it is relatively simple to answer the question of how much
 supersymmetry is preserved on the various fixed-planes of a given
 orbifold, provided the action of the associated quotient group has a well-defined
 lift to the eleven-dimensional spinorial supercharge.  It is therefore a
 straightforward exercise to categorize a wide class of supersymmetric
 orbifolds in {\it M}-theory.  Presumably these singular constructions can be
 resolved to smooth, compact $G_2$ manifolds
 \cite{joyce}, and therefore
 provide a skeletal basis for the characterization of such spaces.
 Especially useful are the stringent
 constraints, based on local anomaly cancellation, which allow one to readily
 discern chiral states and additional characteristic classes (``rational bundle data")
 localized on the
 network of fixed planes.  In this way, many {\it M}-theory orbifold models are very
 similar to string theory brane-world constructions.

 In the interest of developing a robust and useful algorithm for extracting
 effective physics from generic supersymmetric {\it M}-theory orbifolds,
 we analyze and resolve the network of
 constraints which follow from the necessary requirement of
 gauge and gravitational anomaly cancellation point-wise in
 eleven dimensions. Quite a bit of the necessary apparatus has been
 developed in preceding papers \cite{4d1, mlo, phase, hetk3}.
 With some care the technology described in those papers can be
 applied to a wide class of models.  It is opportune,
 therefore, to investigate which orbifold constructions
 are especially interesting, so that we can proceed to cycle through
 these, model-by-model, in the interest of identifying those
 which have the greatest phenomenological appeal, to enable an
 appropriately thorough comparison with string theory models,
 and to learn as much as possible about the world of
 {\it M}-theory physics.

 Considering {\it M}-theory on a spacetime with topology
 $\R^{3,1}\times X^7$, the preservation of $N=1$
 supersymmetry requires that $X^7$ have $G_2$ holonomy \cite{Ach}.
 Furthermore, the presence of chiral fermions and non-abelian gauge
 symmetries in four dimensions adds another requirement to the
 structure of $X^7$, namely, this space cannot be smooth;
 it must possess singularities of one sort or another \cite{AtWit,CandR}.
 One constructive approach, which easily includes both the
 $G_2$ requirement and also the requirement of singularities
 is to focus on global orbifold constructions.  In this case,
 we can replace the geometric holonomy constraint with
 the requirement that some components of the
 eleven-dimensional supercharge are preserved at each point.
 Since this can be readily implemented on global
 orbifolds $T^7/\Gamma$, rather than on merely local models
 of orbifold singularities (e.g., $\R^7/\Gamma$), we gain insight
 into the physics corresponding to global $G_2$
 compactifications.

 More generally, we would like to study all possible
 orbifolds $T^7/\Gamma$, where the torus $T^7$ is itself defined as
 a quotient $\R^7/\Lambda$, with $\Lambda$ a generic lattice in
 $\R^7$ and $\Gamma$ a subset of the automorphisms of $\Lambda$.
 It is worthwhile, however, to restrict attention to an
 important subset of these constructions, namely those
 for which $\Gamma\subset SO(7)\subset SO(10,1)$ is an abelian
 finite group represented by rotations in three complex planes
 plus the possibility of a parity flip in an additional
 real coordinate.  In these cases each element of $\Gamma$
 lifts to an action on spinors, such as the supercharge $Q$,
 in a manner which is especially amenable to analysis.
 One can thereby readily determine the set of supersymmetric
 orbifolds of this class.

 We make one more important restriction: we limit
 attention to those orbifolds for which no group element
 acts freely on any of the coordinates
 of $T^7$.  We call these {\it hard} orbifolds.
 In this case, each element has fixed
 planes associated with it, and the set of fixed
 planes generically intersect as an intricate tangle.
 Geometrically, these models are
 more interesting than the related cases in which
 $\Gamma$ includes elements with fixed-point-free ``shifts" on one or more coordinates.
 These latter types we call {\it soft} orbifolds.
 One can consider the hard orbifolds as more fundamental, because each soft
 orbifold can be obtained from a hard orbifold by
 ``softening" operations in which, for instance, a coordinate
 reflection is
 replaced with a shift.  Geometrically, such softening operations
 either eliminate some of the fixed planes, or they take two fixed planes
 which intersect, and move them off of each other, thereby eliminating
 the intersection.

 The number of hard abelian orbifold models which maintain supersymmetry is
 surprisingly restrictive.  For instance, if one looks only at
 finite groups $\Gamma$ with order less than or equal to
 twelve, there are exactly $29$ such models, as depicted in
 Table \ref{scan}.
\begin{table}
 \begin{center}
 \begin{tabular}{|c|c|ccccccc|}
 \hline
 &
 & \hspace{.3in}
 & \hspace{.3in}
 & \hspace{.3in}
 & \hspace{.3in}
 & \hspace{.3in}
 & \hspace{.3in}
 & \hspace{.3in}  \\[-.1in]
 $\Gamma$ & ${\rm Ord}(\Gamma)$ & 1 & 2 & 3 & 4 & 5 & 6 & 7 \\[.1in]
 \hline
 &&&&&&&&\\[-.1in]
 $\Z_2$               & 2 & $1^*$ & & & 1 & 1 &  &  \\[.1in]
 $\Z_3$               & 3 &       & & & 1 &  &  &  \\[.1in]
 $\Z_4$               & 4 &       & & & 1 & 1 & 1 & 1  \\[.1in]
 $\Z_2\times \Z_2$    & 4 &       & & &   & $1^*$ & 1 & 1 \\[.1in]
 $\Z_2\times \Z_3$    & 6 &       & & & 1 & $1^*$+1 & 1 & 1 \\[.1in]
 $\Z_2\times \Z_4$    & 8 &       & & &   & $1^*$ & $1^*$+1 & $1^*$+2 \\[.1in]
 $(\Z_2)^3$           & 8 &       & & &   &  &  & $1^*$+1 \\[.1in]
 $(\Z_3)^2$           & 9 &       & & &   &  &  &  \\[.1in]
 $(\Z_2)^2\times \Z_3$& 12&       & & &   & $1^*$ & 1 & $1^*$+2 \\[.1in]
 $\Z_3\times \Z_4$    & 12&       & & &   &  &  &  \\[.1in]
 \hline
 \end{tabular} \\[.3in]
 \caption{A scan of supersymmetric hard abelian orbifolds of
 {\it M}-theory.  Non-zero numbers in the table indicate the
 multiplicity of supersymmetric orbifolds $T^n/\Gamma$
 for the indicated abelian finite groups.
 The torus dimension $n$ corresponds to the column and the quotient
 group $\Gamma$ correlates with the row.  Stars indicate supersymmetric
 Horava-Witten models: those which include an isolated $S^1/\Z_2$ factor.
 Note that there are nine of these in the scan, six of which have been
 described previously; the simplest of the remaining three, namely
 the starred $T^7/(\Z_2\times\Z_2\times\Z_3)$ model, is described in this paper.}
 \label{scan}
 \end{center}
 \end{table}
 Table \ref{scan} indicates each supersymmetric
 hard abelian orbifold ${\cal O}=T^7/\Gamma$ with ${\rm Order}(\G)\le 12$.
 In general, the group $\Gamma$ might act non-trivially on a subset of
 the seven coordinates of $T^7$, so that we could
 write instead ${\cal O}=(S^1)^{7-n}\times T^n/\Gamma$.
 Thus, we separate the cases with distinct values of $n$,
 and list these in separate columns in our scan.
 The numbers which appear in the scan are the multiplicities
 of supersymmetric models $T^n/\Gamma$ which can be formed
 by representations of $\Gamma$ on an $n$-torus
 defined by a particular compatible lattice.

 An especially interesting
 subset of the supersymmetric orbifold models
 are those which split off a separate
 $S^1/\Z_2$ factor, since these models have fixed ten-planes.
 We refer to such models as Ho{\v r}ava-Witten, or HW, models, since
 the most basic of these cases was first described in
 \cite{hw1, hw2}.
 These are interesting because local gravitational anomaly cancellation
 requires ten-dimensional $E_8$ Yang-Mills multiplets on these
 ten-planes.  As it turns out, when these ten-planes intersect
 other fixed-planes, further anomaly cancellation requirements
 are satisfied only if the quotient group $\Gamma$ acts
 non-trivially on the $E_8$ lattices, thereby breaking these
 groups down to subgroups.  This allows concise analytical access to information
 pertaining to localized rational bundle data, related to small
 instantons stuck on fixed-plane intersections.

 In Table \ref{scan} we have indicated the HW models
 with an asterix.  Note that there are exactly nine
 supersymmetric hard abelian HW models with
 ${\rm Order}(\Gamma)\le 12$.
 The first is
 the basic $S^1/\Z_2$ model described in \cite{hw1, hw2}.
 Next are the four global orbifold limits of
 $K3\times S^1/\Z_2$, which were analyzed in
 \cite{mlo, phase, hetk3, ksty}.
 Finally, there is one five-dimensional model, wherein only six of the $T^7$ coordinates are
 influenced nontrivially by $\Gamma$, and three four-dimensional models, wherein
 all of the $T^7$ coordinates are influenced nontrivially by
 $\Gamma$.  One of the four-dimensional models has
 $\Gamma=(\Z_2)^3$, and was described in \cite{4d1}.
 In that model, however, the four-dimensional effective
 physics is not chiral, a circumstance related to the fact that
 all the elements of the orbifold group have order two.
 Furthermore, that model does not have purely four-dimensional
 fixed-planes associated with any of the elements of $\Gamma$.
 For this reason, that model does not describe a true four-dimensional
 brane-world.  Of the two remaining four-dimensional models, one has a quotient
 group with prime factors, and one has non-prime factors.
 The latter of these, corresponding to $T^7/(\Z_2\times \Z_4)$
 is relatively complicated.  This is because in that case, in addition to
 a primary category of orbifold-planes, there is a separate
 subclass of hyperplanes comprising nontrivial multiplets under
 the subgroup $(\Z_2)^2$.  Such matters, pertaining to non-prime
 orbifolds were explained more comprehensively in the context of
 $T^5/\Gamma$ orbifolds in \cite{hetk3}.
 The one remaining four-dimensional HW model in our scan has
 $\Gamma=(\Z_2)^2\times\Z_3$.  This model has both
 four-dimensional fixed planes and also a chiral four-dimensional
 spectrum.  Thus, this model is unique in that it is the
 simplest hard abelian orbifold with ten-dimensional fixed-planes,
 giving rise to a chiral four-dimensional super Yang-Mills theory.

 In the bulk of this paper, we provide a microscopic anomaly
 analysis on the unique HW orbifold $T^7/(\Z_2\times\Z_2\times\Z_3)$
 described above.  Our motivation for explaining this model in
 detail has less to do with the particulars of the associated four-dimensional physics
 than it has to do with exposing the set of techniques
 which we employ.  Specifically, this analysis
 rounds out the analytical tools developed in
 \cite{4d1, mlo, phase, hetk3}, filling in the final
 part of the technology: the analysis of the four
 dimensional gauge and mixed anomalies induced at four-dimensional
 fixed planes.

 An interesting feature of hard orbifold models is the way in
 which the perceived four-dimensional
 gauge groups are embedded within larger groups localized on higher-dimensional
 planes.  This in turn is governed by entwined branchings which correlate
 with the manner in which various orbifold planes intersect.
 Different branchings in this context correspond to different
 classes of small instantons living on the branes.
 In the analysis described in this paper, we do not include
 fivebranes.  Therefore, we are
 describing a ``basic" solution, from which additional models
 can be built up by the sorts of phase transitions described
 in \cite{phase}.  This paper is structured as follows.

 In section 2 we describe in detail the construction of
 the particular $T^7/(\Z_2\times \Z_2\times \Z_3)$ orbifold
 described above. We exhibit the
 representation of the quotient group on the compact coordinates,
 and then characterize the intricate geometry of the
 intersecting hyper-planes invariant
 under elements of this group.  In section 3 we explicitly
 derive the spectrum of states localized on each of the orbifold
 fixed-planes described in section 2.  This analysis relies
 on local anomaly cancellation on each ten-, six- and
 four-dimensional fixed-plane in the orbifold, and involves
 the notion of ``consistently entwined branchings", which we
 define and describe.  In section 4, we use the results of
 section 3 to determine the effective spectrum associated with
 complementary four-dimensional brane-worlds linked by a
 five-dimensional bulk, obtained by taking a ``spindle" limit
 in which six of the compact dimensions become small.  We then
 conclude with various observations about the relationship of
 {\it M}-theory models, such as the one described in this paper,
 with analogous constructions derived from within string theory.

 \section{Fixed-Plane Geometry}
 The simplest supersymmetric hard global orbifold
 \footnote{As explained in the introduction,
 a hard orbifold is defined as one with a quotient
 group which has no elements which act freely on any coordinate.}
 of {\it M}-theory which has four-dimensional fixed-planes and
 a chiral four dimensional spectrum has the following structure.
 The eleven-dimensional spacetime has topology
 $\R^{3,1}\times T^6/(\Z_2\times \Z_3)\times S^1/\Z_2$.
 The six-torus is defined as a lattice quotient $\R^6/\Lambda$,
 where $\Lambda=A_2\oplus A_2\oplus A_1$, and is parameterized
 by three complex coordinates $(\,z_1\,,\,z_2\,,\,z_3\,)$
 \footnote{The lattice in question is
 defined by the identifications $z_i\to z_i+1$ and $z_i\to \exp{(\,2\,\pi\,i\,l_i\,)}$
 where $l_i=(1/3,1/3,1/4)$.  Thus, $\Lambda$ is a direct
 sum of two hexagonal lattices and one square lattice.}.
 The circle $S^1$ is described by a real angular coordinate $x^{11}$.
 The quotient group $\Z_2\times\Z_2\times \Z_3$
 acts on the coordinates as indicated in Table \ref{rules}.
 \begin{table}
 \begin{center}
 \begin{tabular}{c|cccc}
 & $z_1$ & $z_2$ & $z_3$ & $x^{11}$ \\[.1in]
 \hline
 &&&&\\[-.1in]
 $\a$ & $-$ & + & $-$ & + \\[.1in]
 $\b$ & + & + & + & $-$ \\[.1in]
 $\a\b$ & $-$ & + & $-$ & $-$ \\[.1in]
 $\g$ & 1/3 & $-1/3$ & + & + \\[.1in]
 $\a\g$ & $-1/6$ & $-1/3$ & $-$ & + \\[.1in]
 $\b\g$ & 1/3 & $-1/3$ & + & $-$ \\[.1in]
 $\a\b\g$ & $-1/6$ & $-1/3$ & $-$ & $-$ \\
 \multicolumn{5}{}{} \\[1.6in]
 \end{tabular}
 \vspace{-1.7in}
 \caption{The representation of $\Gamma=\Z_2\times\Z_2\times\Z_3$
 on the coordinates of the seven torus
 $(\,z_1\,,\,z_2\,,\,z_3\,,\,x^{11}\,)$, for the orbifold described
 in the text. This order twelve group is generated by the elements $\a, \b$ and $\g$.
 A minus sign on a complex coordinate $z_i$ is equivalent to an $f_i=1/2$ rotation.
 In the table we have suppressed the trivial element and the
 inverses of the order 3 element $\g$ and the three order 6 elements $\a\g$,
 $\b\g$ and $\a\b\g$. }
 \label{rules}
 \end{center}
 \end{table}
 In this representation, each element acts as a rotation in the three complex planes and
 possibly a parity flip in the $S^1$ direction,
 \brr (\,z_1\,,\,z_2\,,\,z_3\,,\,x^{11}\,) \rightarrow
      (\,e^{i\,\q_1} z_1\,,\,e^{i\,\q_2} z_2\,,\, e^{i\,\q_3} z_3\,,\,(-)^P x^{11}\,) \,.
 \err
 with $\q_i = 2\,\pi\,f_i$ and $P\in \{0,1\}$.
 Depicted in Table \ref{rules} are the fractional rotations $f_i$
 imparted on the planes and the presence or absence of an $x^{11}$
 parity flip, for each representative non-trivial element of the group.
 Note that the four elements $\{\,\g^2\,,\a\g^2\,,\,\b\g^2\,,\,\a\b\g^2\,\}$
 are the respective inverses of the four elements
 $\{\,\g\,,\a\g\,,\,\b\g\,,\,\a\b\g\,\}$,
 and have precisely the same fixed planes.
 We have therefore suppressed four nontrivial elements
 in Table \ref{rules}.

 The element $\b$ has two ten-dimensional fixed planes: an
 ``upper" one and a ``lower" one, each corresponding to a
 separate value of $x^{11}$.  The other fixed-planes
 (i.e. those associated with other elements of $\Gamma$)
 fall into two categories: a primary set
 comprised of subspaces of the $\b$-invariant ten-planes, and a
 secondary set involving those which span $x^{11}$.  Each of the
 secondary fixed-planes
 interpolates between pairs of primary fixed planes.
 The primary fixed-planes are associated with the elements
 $\a\b$, $\b\g$ and $\a\b\g$; those associated with
 $\a\b$ and $\b\g$ are six-dimensional, while those associated
 with $\a\b\g$ are four-dimensional and coincide with
 intersections of the six-dimensional primary fixed-planes.
 The secondary fixed
 planes are associated with the elements $\a$, $\g$ and $\a\g$;
 those associated with $\a$ and $\g$ are
 seven-dimensional, while those associated with $\a\g$ are five-dimensional
 and coincide with intersections of the seven-dimensional
 secondary fixed planes.
 The $\a$-invariant seven-planes interpolate between pairs of
 $\a\b$-invariant six-planes,
 the $\g$-invariant seven-planes interpolate between pairs of $\b\g$-invariant
 six-planes, and the $\a\g$-invariant five-planes interpolate between
 $\a\b\g$-invariant four-planes.

 We first analyze the geometry of the secondary fixed-planes
 (i.e. those which span $x^{11}$).
 We are therefore interested in studying the action of
 $\a$, $\g$ and $\a\g$ on the three complex
 coordinates $(\,z_1\,,\,z_2\,,\,z_3\,)$, at generic values
 of $x^{11}$.  The $\g$-planes are the only
 secondary planes which span the $z_3$ directions.
 The element $\g$ has order-three and acts on
 $(z_1,z_2)$, providing a set of nine ostensibly isolated $A_2$ orbifold
 singularities, each of the sort characteristic of a $\C^2/Z_3$ orbifold.
 However, at four special values of $z_3$ the elements
 $\a$ and $\a\g$ identify pairs of points within this set,
 effectively inducing intersections.
 At generic values of
 $z_3$ the secondary planes consist exclusively of
 nine $\g$-invariant seven-planes. At the four special values of
 $z_3$, however, the geometry is comparatively intricate.

 We focus on subspaces $T^4\subset T^6$ which are
 spanned by $(z_1,z_2)$, at the four special values of $z_3$.
 The coordinates $z_1$ and $z_2$ each take values in
 a fundamental domain of an $A_2$ lattice.
 It is useful first to consider one such domain, that associated
 with $z_1$, which we
 depict as follows,
 \begin{center}
 \includegraphics[width=1.5in,angle=0]{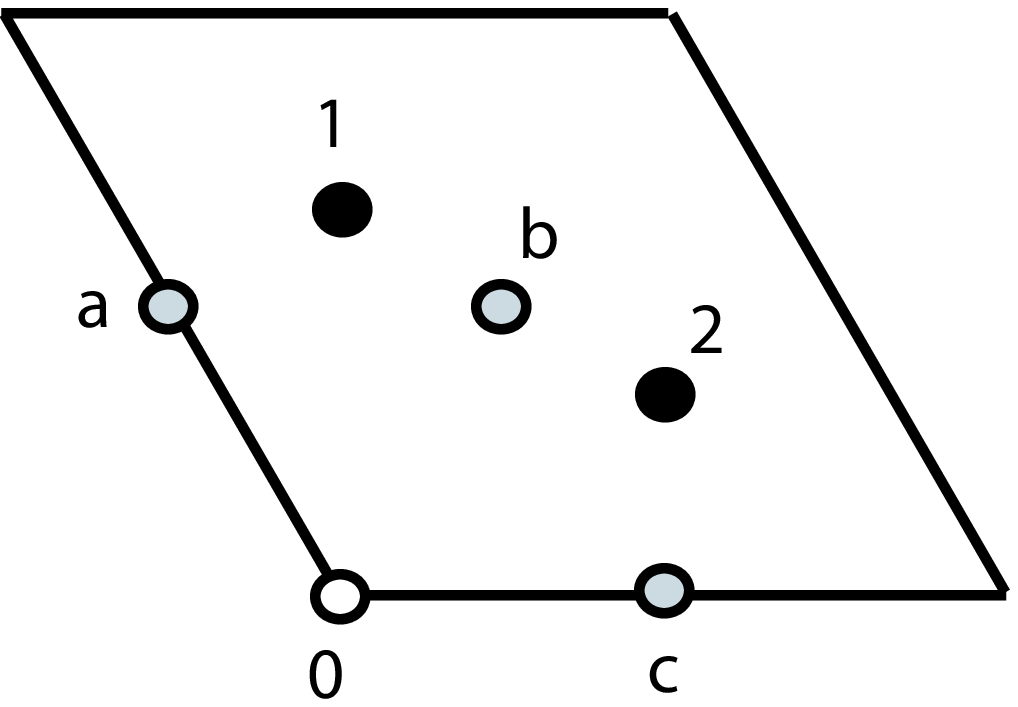}\\[.2in]
 \end{center}
 Here we have indicated special points which are fixed
 under relevant $\Z_2$, $\Z_3$ or $\Z_6$ actions
 generated by $\a$, $\g$ and $\a\g$, respectively.  The origin
 of this complex plane is denoted $0$.  The two points
 labelled $1$ and $2$ are $\Z_3$ invariants, but transform
 as doublets under $\Z_2$.  The three points labelled
 $a$, $b$ and $c$ are $\Z_2$ invariants, but transform
 as a triplet under $\Z_3$.  The point labelled $0$
 is invariant under both $\Z_2$ and $\Z_3$ and is the only point
 invariant under $\Z_6$.  Now consider the fundamental
 domain of the second $A_2$ lattice, that spanned by $z_2$.
 In this case, we have a picture similar to that described above,
 but with a crucial difference: the element $\a$ does not act on $z_2$.
 Thus, whereas the first $A_2$ lattice had only
 four $\a$-invariant points, 0, $a$, $b$ and $c$, the second $A_2$ lattice
 is $\a$-invariant in its entirety.

\begin{figure}
 \begin{center}
 \includegraphics[width=2.5in,angle=0]{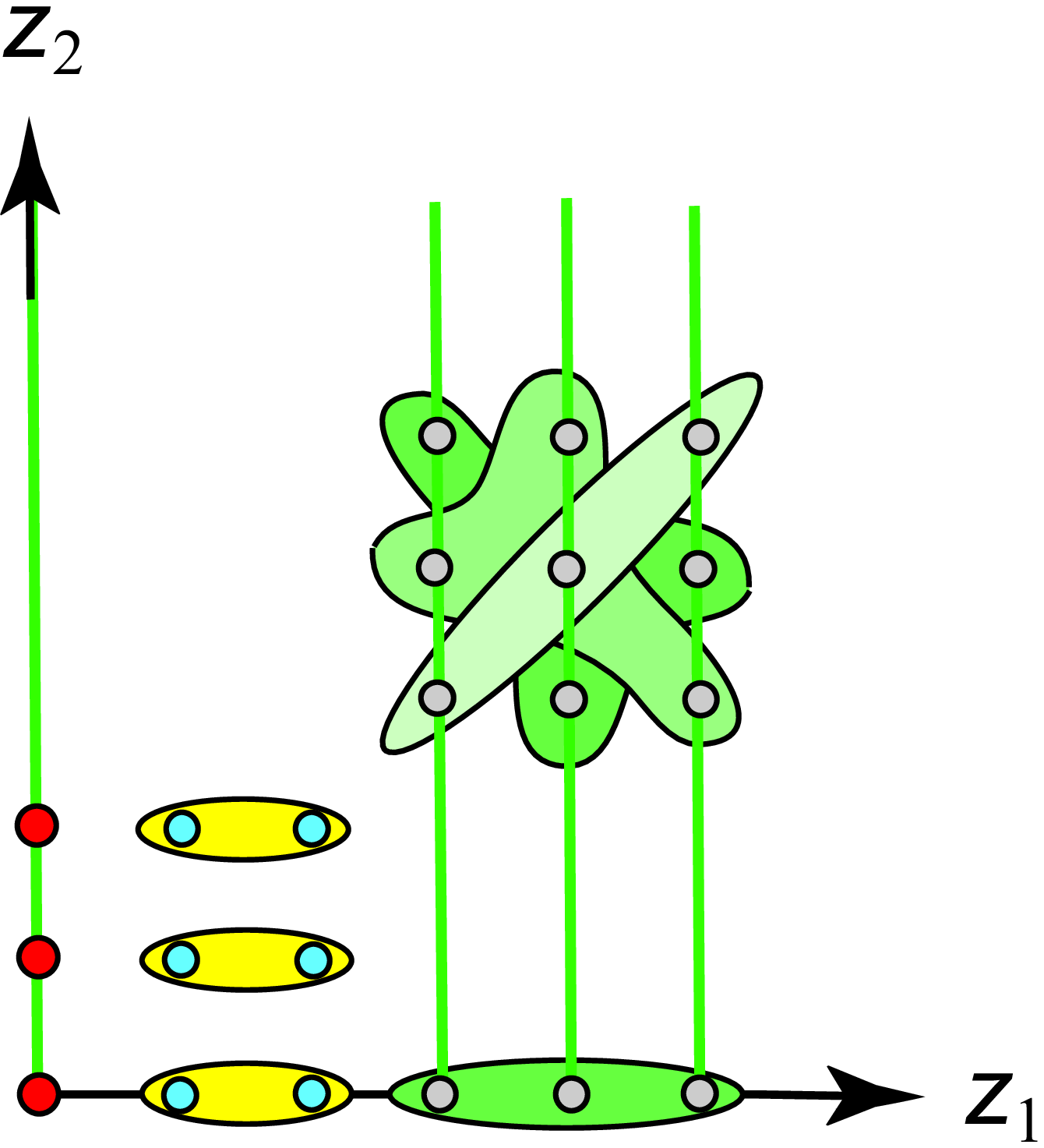}\\[.2in]
 \caption{Points of the four-torus $(z_1,z_2)$ identified
 by the $\Z_2\times\Z_3$ subgroup of $\Gamma$ generated by $\a$ and $\g$.
 The red spots are the three $\Z_2\times\Z_3$ invariant points,
 the yellow blobs are the $\Z_3$-invariant
 $\Z_2$-doublets with components in blue.  The
 green lines are the four $\Z_2$ invariants.  The green blobs
 encircle those $\Z_3$ triplets which provide noteworthy
 identifications.}
 \label{torid}
 \end{center}
 \end{figure}

 Special points in the combined $A_2\oplus A_2$
 lattice parameterized by $(z_1,z_2)$ are represented
 in the obvious manner by pairs, such as $(0,0)$ or $(a,2)$.
 Within $(z_1,z_2)$, there are four parallel codimension two loci
 invariant under $\a$.  These are given by $(0,z_2)$, $(a,z_2)$,
 $(b,z_2)$ and $(c,z_2)$, and are depicted by the green lines
 in Figure \ref{torid}.  Next, there are nine points
 invariant under $\g$.  Three of these are given by $(0,0)$,
 $(0,1)$ and $(0,2)$, and are depicted in red in
 Figure \ref{torid}. The other six, which comprise three
 doublets under $\a$, are given by
 $(1,0)\leftrightarrow (2,0)$, $(1,1)\leftrightarrow (2,1)$
 and $(1,2)\leftrightarrow (2,2)$,
 where the arrows indicate the $\Z_2$ transformations
 generated by $\a$.
 (In Figure \ref{torid}, these six points are depicted in blue, while the
 identifications corresponding to the order-two element
 $\a$ are indicated by the yellow blobs.)
 There are also several noteworthy $\Z_3$ identifications,
 as indicated in Figure \ref{torid} by the green blobs
 encircling triplets of grey points.
 For example, the three points $(0,a)$, $(0,b)$ and $(0,c)$
 comprise one such $\Z_3$ triplet.
 Thus, at the four special values of $z_3$, the geometry
 inside the $T^4$ parameterized by $(z_1,z_2)$ includes
 three $\Z_6$-invariant points (red) linked by a $\Z_2$-invariant
 complex line (green), three isolated $\Z_3$-invariant points
 (yellow) and four more $\Z_2$-invariant points (grey) linked by triple
 intersections of $\Z_2$-invariant lines (green).

\begin{figure}
 \begin{center}
 \includegraphics[width=4.3in,angle=0]{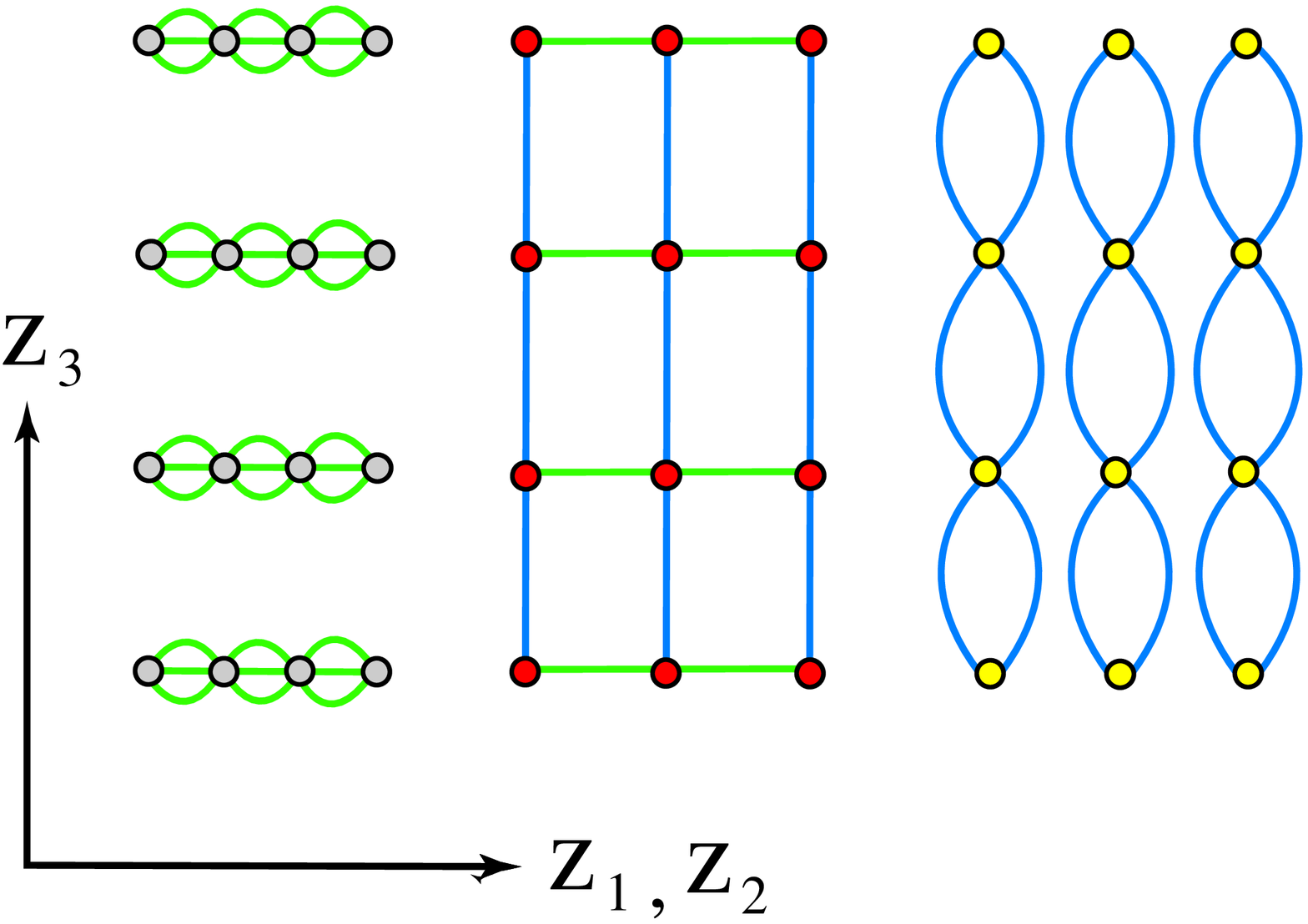}\\[.2in]
 \caption{Fixed loci inside one of the six-tori,
 parameterized by $(z_1,z_2,z_3)$,
 at one end-of-the world (i.e. at one of the two special values of
 $x^{11}$).
 The red points are the twelve
 $\a\b\g$-invariant four-planes, the blue lines are the nine $\b\g$-invariant
 six-planes, and the green lines are the sixteen $\a\b$-invariant
 six-planes.  Three of the (blue) $\b\g$-invariant six-planes
 intersect four of the (green) $\a\b$-invariant
 six-planes at (red) $\a\b\g$-invariant four-planes.
 There are twelve such intersections.
 Otherwise, the six remaining (blue) $\b\g$-invariant six-planes
 doubly intersect at the nine four-planes shown
 in yellow, and the twelve remaining (green) $\a\b$-planes
 triply-intersect at the sixteen four-planes shown in grey.
 Six-tori at generic values of $x^{11}$ have a similar geometry.}
 \label{endtori}
 \end{center}
 \end{figure}

 Now lets consider the $z_3$ dependence as well, and describe the
 geometry inside the $T^6/(\Z_2\times \Z_3)$ at a given value of
 $x^{11}$.  Consider, for example, the geometry at one of the
 two special values of $x^{11}$ fixed by the element $\b$, i.e. at one
 ``end-of-the-world".  This can be depicted as shown in
 Figure \ref{endtori}.
 As described previously, the only fixed planes which span $z_3$
 are the $\b\g$-invariant planes, of which there are nine.  These are represented
 by the nine blue lines in Figure \ref{endtori}. At the
 four special values of $z_3$, these are identified pairwise as
 indicated.

 Now we can describe the global geometry associated with this
 orbifold.  At each end of the world we have the network of
 fixed planes shown in Figure \ref{endtori}.  We have a similar
 geometry at generic values of $x^{11}$.  We observe
 three essential sorts of extended neighborhoods.
 First, there are the $\Z_6$-invariant planes
 which each live at the intersection of one $\Z_2$-invariant plane
 and one $\Z_3$ invariant plane.  These correspond to the red dots
 in Figure \ref{endtori}.  Second, there are the triple
 intersections of the $\Z_2$-invariant planes, depicted in grey
 in Figure \ref{endtori}.  Third, there
 are double intersections of $\Z_3$-invariant planes, depicted in
 yellow in Figure \ref{endtori}.

\begin{figure}
 \begin{center}
 \includegraphics[width=2.2in,angle=0]{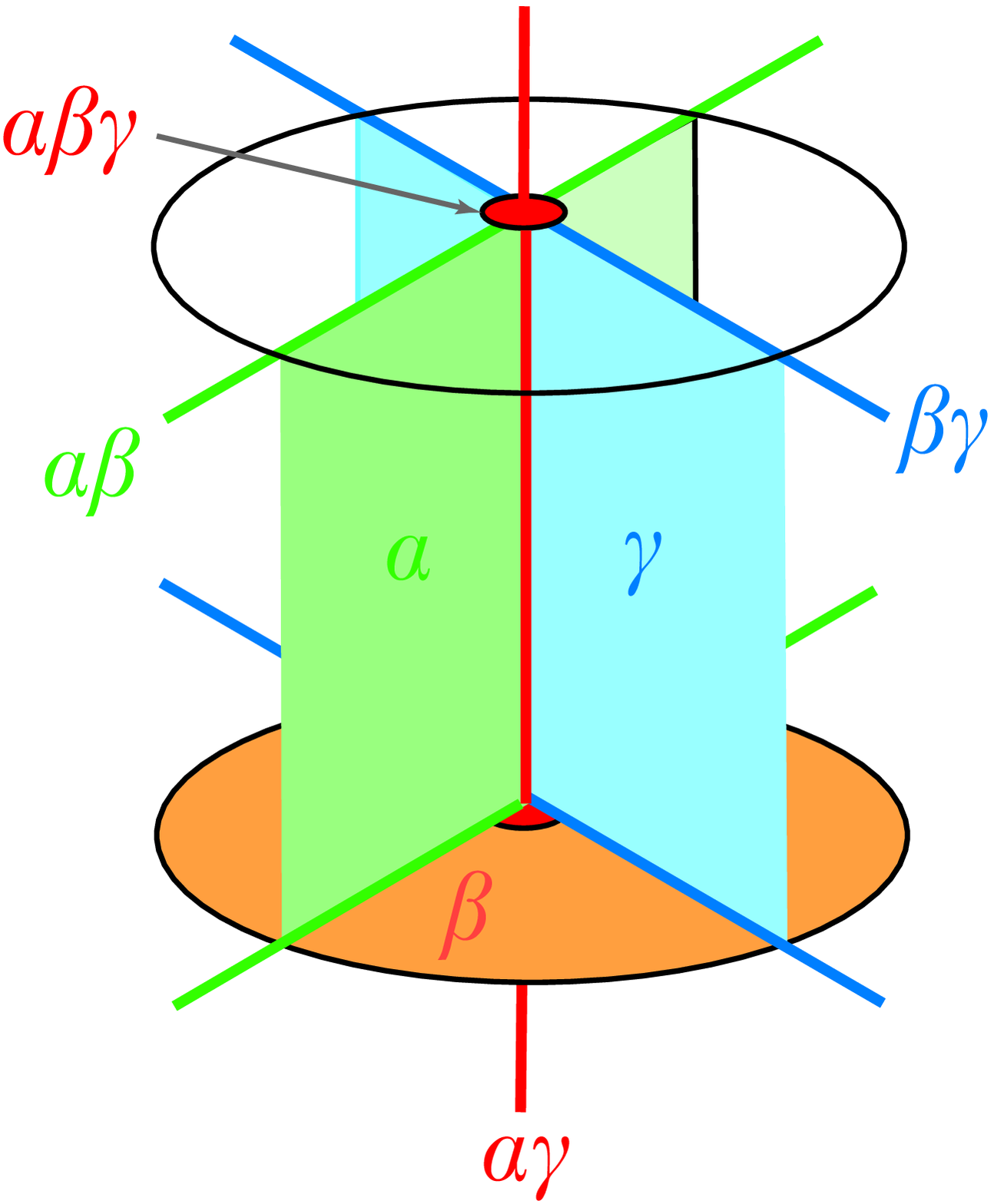} \\[.8in]
 \includegraphics[width=2.5in,angle=0]{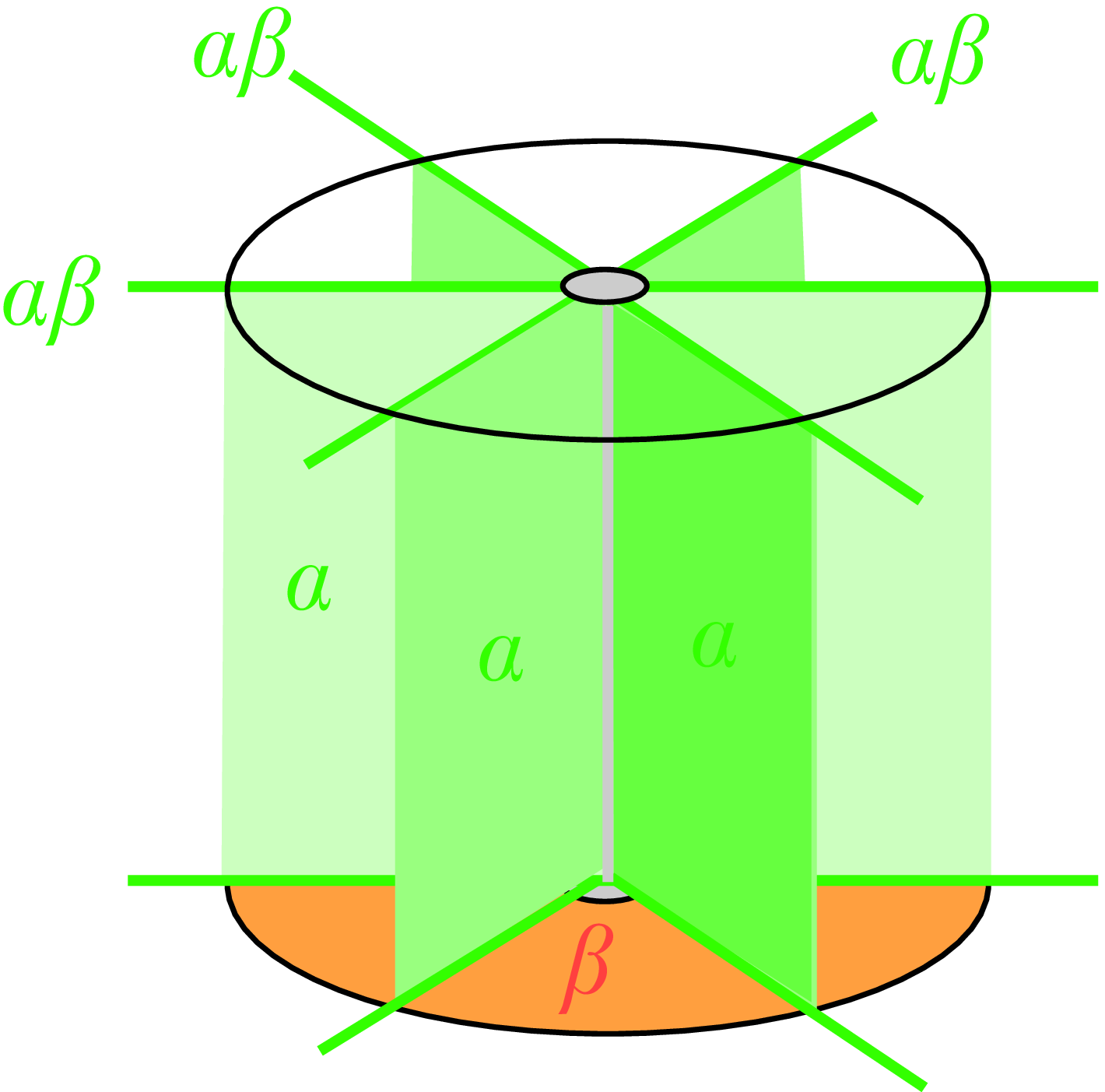}
 \hspace{1in}
 \includegraphics[width=2.2in,angle=0]{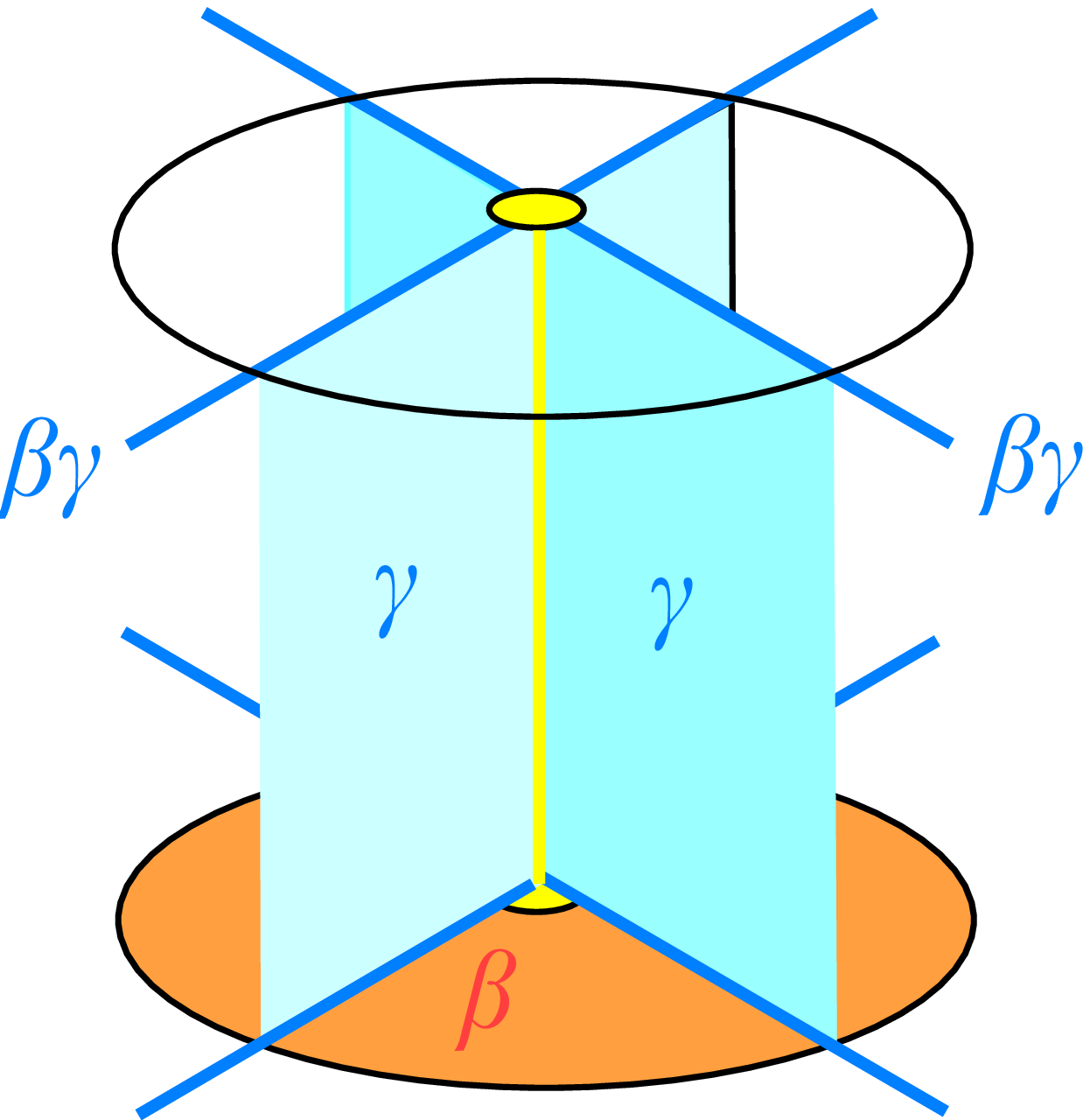}
 \\[.3in]
 \caption{A local depiction showing each of the three sorts of
 extended neighborhoods characteristic of the orbifold described in the text.
 In the first instance, we exhibit two of the twelve $\a\b$-invariant
 four-planes (red circles) connected by one of the $\a\g$-invariant
 five-planes (the red line). This is an extended neighborhood of the
 first category, as explained in the text.  The lower two pictures
 depict extended neighborhoods of the second and third categories.}
 \label{wheels}
 \end{center}
 \end{figure}

 Neighborhoods of the first category extend from one
 $\a\b\g$-invariant four-plane located at the intersection
 of an $\a\b$-invariant six-plane and a $\b\g$-invariant
 six-plane, all within one end-of-the-world (i.e. all within one
 $\b$-invariant ten-plane), to a second one within the other end-of-the-world.
 The plane which interpolates between the two $\a\b\g$-invariant four-planes
 is an $\a\g$-invariant five-plane, which lives at the
 intersection of an $\a$-invariant seven-plane and a
 $\g$-invariant seven-plane.  One such extended neighborhood is
 depicted in the uppermost picture in Figure \ref{wheels}.
 There are twelve extended neighborhoods of the first category in
 this orbifold.

 Neighborhoods of the second category extend from triple
 intersections, each involving three $\a\b$-invariant six-planes within
 one end-of-the-world, to similar triple intersections within the
 other end-of-the-world.  The planes which interpolate, spanning $x^{11}$,
 between pairs of the $\a\b$-planes are themselves $\a$-invariant
 seven-planes.  In this case the four-dimensional intersections
 and the five-dimensional interpolating intersection are not
 by-themselves invariant under any elements of the quotient group.
 Instead, these intersections comprise triplets under the $\Z_3$
 generated by $\g$.   One such extended neighborhood is depicted
 in the lower left picture in Figure \ref{wheels}.  There are
 sixteen extended neighborhoods of the second category in this
 orbifold.

 Neighborhoods of the third category extend from double
 intersections, each involving two $\a\g$-invariant six-planes
 within one end-of-the-world, to a similar double intersection
 within the other end-of-the-world.  The planes which
 interpolate, spanning $x^{11}$, between pairs of the $\a\g$-planes
 are themselves $\g$-invariant seven-planes.  In this case the
 four-dimensional intersections and the five-dimensional
 interpolating intersection are not by-themselves invariant under any
 elements of the quotient group.  Instead, these intersections
 are triplets under the $\Z_2$ generated by $\a$.  One such
 extended neighborhood is depicted in the lower right picture
 in Figure \ref{wheels}.  There are twelve extended
 neighborhoods of the third category in this orbifold.

 \section{Localized States}
 Now that we have characterized the network of fixed planes
 in our orbifold, we address the issue of potential chiral anomalies
 localized on these planes.

 The bulk gravitino field is
 projected chirally by the element $\b$ onto the the
 $\b$-invariant ten-planes defining the ends-of-the-world.
 The chiral coupling of this bulk field to currents localized on
 these ten-planes induces a localized gravitational anomaly.  This is
 eliminated self-consistently (i.e. avoiding the introduction of additional
 gauge or mixed anonalies) by including $E_8$ Yang-Mills
 super-multiplets on each ten-plane.  However, the elements
 $\a\b$ and $\b\g$ also act on the bulk gravitino field, and on
 the $E_8$ gaugino fields as well, so as to introduce additional
 gravitational, gauge and mixed anomalies on the six-planes
 associated with $\a\b$ and $\b\g$.  These can also be
 eliminated self-consistently by including $SU_2$ and
 $SU_3$ Yang-Mills super multiplets, on the {\it seven}-dimensional
 $\a$ and $\g$ planes, respectively, and by adding onto the six-planes
 additional ``twisted" hypermultiplets transforming in particular
 representations.  But this is possible only
 if we include as well additional electric and magnetic couplings, and
 only if we impose that $\a$ and $\g$ act nontrivially on the
 $E_8$ gauge lattices.  In the absence of four-dimensional
 intersections, these matters can be analyzed precisely as described
 in \cite{mlo, phase, hetk3}.  However, the intersections add interesting
 new consistency requirements.

 The nontrivial action of $\a$ on the $E_8$ gauge lattice
 implies a breakdown $E_8\to {\cal G}_\a$ as one moves
 within one of the $\b$-invariant ten-planes and then lands on one
 of the $\a\b$-invariant six-planes which is a submanifold.
 Note that $\b$ necessarily acts
 trivially on the $E_8$ lattices because the $E_8$ lattices
 themselves are associated with the $\b$-invariant planes.
 Since $\a$ has order-two, there are special limitations on
 which subgroups ${\cal G}_\a\subset E_8$
 are possible.  The possibilities are also constrained by anomaly
 cancellation.  Two alternative possibilities for ${\cal G}_\a$ turn out to be
 $E_7\times SU_2$ and $SO_{16}$.
 Similarly, the nontrivial action of $\g$ on the $E_8$ gauge lattice
 implies a breakdown $E_8\to {\cal G}_\g$ as one moves
 within one of the $\b$-invariant ten-plane and then lands on one
 of the $\b\g$-invariant six-planes which is a submanifold.
 Since $\g$ has order-three, there are again limitations on
 which subgroups ${\cal G}_\g\subset E_8$
 are possible.  The possibilities are also constrained by anomaly
 cancellation.  Two alternative possibilities for ${\cal G}_\g$ turn out to be
 $E_6\times SU_3$ and $SU_9$.

\begin{figure}
 \begin{center}
 \includegraphics[width=2.3in,angle=0]{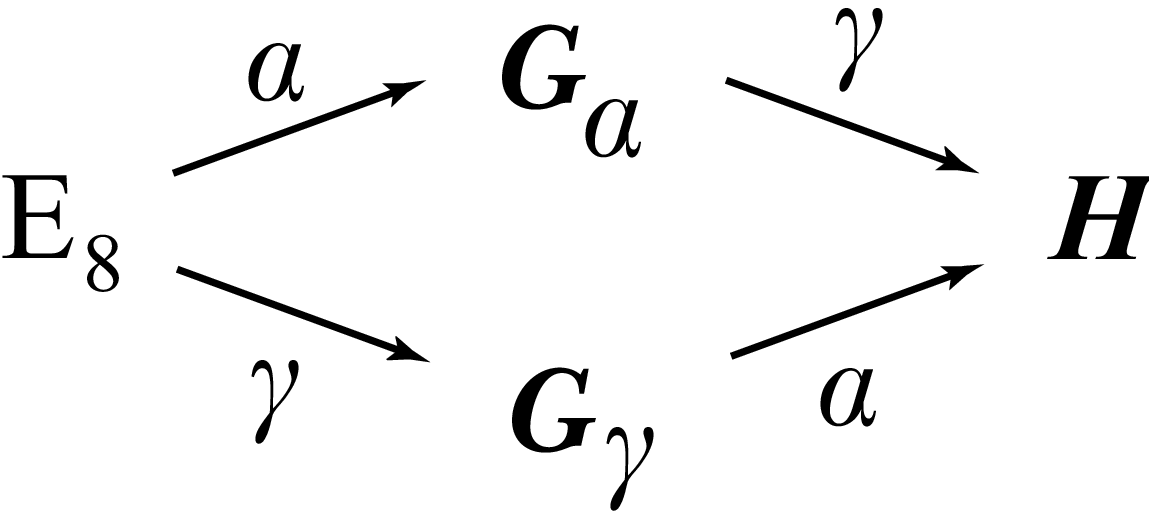}\\[.2in]
 \caption{Entwined branching pattern induced by multiple lattice
 projections at a brane intersection.}
 \label{nesting}
 \end{center}
 \end{figure}

 We focus attention on the $\a\b\g$-invariant four-planes.
 These each live at the intersection of one $\a\b$-invariant six-plane
 and one $\b\g$-invariant six-plane. It is at these intersections that
 extra constraints apply.  If we move within one of the
 $\b$-invariant ten-planes, and then land on one of the
 $\a\b$-invariant six-planes, and {\it then} move within
 this particular six-plane and ultimately land on one of the
 $\a\b\g$-invariant four-planes, we would
 see the $E_8$ group successively broken down according to
 $E_8\to {\cal G}_\a\to {\cal H}$.  The second branching occurs
 because the $\a\b\g$-invariant four-plane is independently invariant under both
 $\a$ and $\g$, and also because $\a$ and $\g$ have independent
 actions on the $E_8$ lattice.  Thus, $\g$ acts nontrivially on
 the sublattice of $E_8$ corresponding to ${\cal G}_\a$.
 This serves to break ${\cal G}_\a$ down to ${\cal H}$ on
 the $\a\b\g$ four-planes.  Now imagine that we move
 within the same original $\b$-invariant ten-plane that we considered above,
 but this time land first within one of the $\b\g$-invariant six-planes and then
 move within this to ultimately land inside the same four-dimensional intersection
 described previously.  Considerations similar to those described above
 apply in this case, except that this time we observe a successive branching
 with a different subgroup at the intermediary step,
 $E_8\to {\cal G}_\g\to {\cal H}$.  Necessarily the subgroup
 ${\cal H}\subset E_8$ is the same subgroup encountered above.
 This implies that $\a$ and $\g$ collectively generate an entwined branching
 to the subgroup ${\cal H}$ as illustrated in Figure \ref{nesting}.
 In what follows we will refer to an entwined branching
 pattern such as the one shown in Figure
 \ref{nesting} using the notation
 $(\,{\cal G}_\a\,,\,{\cal G}_\g\,|\,{\cal H}\,)$.

 As described above, there are twelve complementary pairs of
 $\a\b\g$-invariant four-planes:
 twelve such planes inside the ten-plane at the upper
 end-of-the-world are paired with twelve more inside the lower
 end-of-the-world.  Elements of a pair are linked by
 interpolating $\a\g$-invariant five-planes.
 Furthermore each $\a\b$-invariant six-plane and each $\b\g$-invariant six-plane
 is a source of $G$-flux.  In our usual terminology, we
 say that these planes have associated magnetic charges.
 These are the M-theory analogs of RR charges in string theory.
 Owing to global considerations, which can be
 derived by integrating the closed form $dG$ over five-cycles
 in the compact space, one deduces that the sum of these charges
 must vanish.  This is the M-theory analog of the requirement of
 tadpole infinity cancellation in string theory.  As
 explained in \cite{4d1} this imposes that
 $\a$ and $\g$ must act on the two $E_8$ lattices
 in a complementary fashion.  Thus, if we have a
 $(\,{\cal G}_\a\,,\,{\cal G}_\g\,|\,{\cal H})$ breakdown choice
 on the upper ten-plane, we must choose a fully complementary
 choice (\,${\cal G}_\a'\,,\,{\cal G}_\g'\,|\,{\cal H}'\,)$
 on the lower ten-plane.
 Complimentary in this case means that
 ${\cal G}_\a$ and ${\cal G}_\a'$ must be chosen {\it one-each} from
 the pair of subgroups $E_7\times SU_2$ and $SO_{16}$ and, similarly,
 ${\cal G}_\g$ and ${\cal G}_\g'$ must be chosen {\it one-each} from
 the pair of subgroups $E_6\times SU_3$ and $SU_9$.  The groups
 ${\cal H}$ and ${\cal H}'$ then depend on the
 choices made.  Owing to this requirement, the number of
 global possibilities is severely limited.  Requiring that
 any additional four dimensional gauge or mixed anomalies
 can be eliminated restricts the choices even further.

 Now, keeping all of these considerations in mind, we proceed
 to study the situation at one of the $\a\b\g$-invariant
 intersections, say one in the upper end-of-the-world.
 Once we find a consistently entwined branching which has
 curable four-dimensional intersection
 anomalies, we are then faced with the additional requirement of finding
 a consistently entwined complementary branching, associated
 with an $\a\b\g$-plane in the lower end-of-the-world, which also
 has curable four-dimensional intersection anomalies.
 Ostensibly, there are four possibilities for choosing intermediate
 groups $(\,{\cal G}_\a\,,\,{\cal G}_\g\,)$ from the allowed
 possibilities. These four possibilities consist of two complimentary pairs however.
 The first of these is the choice
 $(\,E_7\times SU_2\,,\,E_6\times SU_3\,)$ which is complimentary
 to $(\,SO_{16}\,,\,SU_9\,)$.  The second is the choice
 $(\,E_7\times SU_2\,,\,SU_9\,)$ which is complimentary to
 $(\,SO_{16}\,,\,E_6\times SU_3\,)$.  We have
 discerned a consistent picture free from four-dimensional
 anomalies only for the second of these two choices.

 We proceed to explain completely the consistent solution to the constraints
 described above.  First we describe the situation at one of the ``upstairs
 vertices" (i.e. one of the $\a\b\g$-invariant intersections
 within the upper end-of-the-world) and then we describe
 the complimentary situation at one of the ``downstairs vertices"
 (i.e. one of the $\a\b\g$-invariant intersections within the
 lower end-of-the-world).  In each case we exhibit the entwined
 branching patterns and identify the entwined subgroups, ${\cal H}$ upstairs
 and ${\cal H}'$ downstairs.  We also describe the chiral spectrum
 which survives the projections to four-dimensions, and explicitly
 demonstrate the absence of four-dimensional anomalies.

 \subsection{Upstairs Vertices}
 The generators $\a$ and $\g$ act on the lattice associated with the
 upstairs $E_8$ gauge factor  according to
 $(\,E_7\times SU_2\,,\,SU_9\,|\,SU_6\times SU_3\times U_1\,)$.
 We verify that the subgroup
 ${\cal H}=SU_6\times SU_3\times U_1$ is consistently entwined
 inside of $E_8$ by $E_7\times SU_2$ and $SU_9$ by using
 two important consistency checks.  First, we verify that $E_8$
 branches to precisely the same representation of ${\cal H}$
 when the branching occurs via each of the two separate routes
 indicated in Figure \ref{nesting}.  Then we verify that there
 are no non-curable four-dimensional anomalies localized at
 the intersection.  When we refer to a ``consistently
 entwined" branching we are indicating that both of these
 criteria are met.

 First, we consider the $(\a,\g)$ branching, under which
 the $E_8$ lattice is projected first via $\a$ to
 ${\cal G}_\a$, and then this subgroup is projected,
 via $\g$, to ${\cal H}$.  For the case at hand, we have
 \brr E_8 &\stackrel{\a}{\longrightarrow}& E_7\times SU_2
      \nonumber\\[.1in]
      &\stackrel{\g}{\longrightarrow}& SU_6\times SU_3\times U_1
      \nonumber\\[.2in]
      {\bf 248} &\stackrel{\a}{\longrightarrow}&
      [\,(\,{\bf 133}\,,\,{\bf 1}\,)\,\oplus\,
      (\,{\bf 1}\,,\,{\bf 3}\,)\,]\,\oplus\,
      (\,{\bf 56}\,,\,{\bf 2}\,)
      \nonumber\\[.1in]
      &\stackrel{\g}{\longrightarrow}&
      [\,(\,{\bf 35}\,,\,{\bf 1}\,)_0\,\oplus\,
      (\,{\bf 1}\,,\,{\bf 8}\,)_0\,\oplus\,
      (\,{\bf 15}\,,\,{\bf \bar{3}}\,)_0\,\oplus\,
      (\,{\bf \bar{15}}\,,\,{\bf 3}\,)_0
      \nonumber\\[.1in]
      & & \oplus\,(\,{\bf 1}\,,\,{\bf 1}\,)_{+6}\,\oplus\,
      (\,{\bf 1}\,,\,{\bf 1}\,)_0\,\oplus\,
      (\,{\bf 1}\,,\,{\bf 1}\,)_{-6}\,]
      \nonumber\\[.1in]
      & & \oplus\,(\,{\bf 20}\,,\,{\bf 1}\,)_{+3}\,\oplus\,
      (\,{\bf 6}\,,\,{\bf 3}\,)_{+3}\,\oplus\,
      (\,{\bf \bar{6}}\,,\,{\bf \bar{3}}\,)_{+3}
      \nonumber\\[.1in]
      & & \oplus\,(\,{\bf 20}\,,\,{\bf 1}\,)_{-3}\,\oplus\,
      (\,{\bf \bar{6}}\,,\,{\bf \bar{3}}\,)_{-3}\,\oplus\,
      (\,{\bf 6}\,,\,{\bf 3}\,)_{-3} \,.
 \label{up1}\err
 where we have chosen a convenient normalization for the $U(1)$ charge.
 As a useful mnemonic, we have placed brackets around those terms
 in the representation sum which correspond to the adjoint of
 ${\cal G}_\a$.  This is useful for determining how the element
 $\a$ acts on the $E_8$ lattice.  Since $\a$ breaks $E_8$
 down to ${\cal G}_\a$, it follows that $\a$ acts trivially on
 those root vectors corresponding to the bracketed
 representations, and acts non-trivially on the representations
 which are not bracketed..

 Next, we consider the $(\g,\a)$ branching, under which
 the $E_8$ lattice is projected first via $\g$ to
 ${\cal G}_\g$, and then this subgroup is projected
 via $\a$ to ${\cal H}$.  For the case at hand, we have
 \brr E_8 &\stackrel{\g}{\longrightarrow}& SU_9
      \nonumber\\[.1in]
      &\stackrel{\a}{\longrightarrow}& SU_6\times SU_3\times U_1
      \nonumber\\[.2in]
      {\bf 248} &\stackrel{\g}{\longrightarrow}&
      [\,{\bf 80}\,]\,\oplus\,{\bf 84}\,\oplus\,{\bf \bar{84}}
      \nonumber\\[.2in]
      &\stackrel{\a}{\longrightarrow}&
      [\,(\,{\bf 35}\,,\,{\bf 1}\,)_0\,\oplus\,
      (\,{\bf 1}\,,\,{\bf 8}\,)_0\,\oplus\,
      (\,{\bf 1}\,,\,{\bf 1}\,)_0\,\oplus\,
      (\,{\bf 6}\,,\,{\bf 3}\,)_{+3}\,\oplus\,
      (\,{\bf \bar{6}}\,,\,{\bf \bar{3}}\,)_{-3}\,]
      \nonumber\\[.1in]
      & & \oplus\,
      (\,{\bf 20}\,,\,{\bf 1}\,)_{+3}\,\oplus\,
      (\,{\bf 15}\,,\,{\bf \bar{3}}\,)_0\,\oplus\,
      (\,{\bf 6}\,,\,{\bf 3}\,)_{-3}\,\oplus\,
      (\,{\bf 1}\,,\,{\bf 1}\,)_{-6}
      \nonumber\\[.1in]
      & & \oplus\,
      (\,{\bf 20}\,,\,{\bf 1}\,)_{-3}\,\oplus\,
      (\,{\bf \bar{15}}\,,\,{\bf 3}\,)_0\,\oplus\,
      (\,{\bf \bar{6}}\,,\,{\bf \bar{3}}\,)_{+3}\,\oplus\,
      (\,{\bf 1}\,,\,{\bf 1}\,)_{+6}
  \label{up2}\err
  We have enclosed with
  brackets those terms in the representation sum corresponding
  to the adjoint of ${\cal G}_\g$.  Not surprisingly, these
  are not the same terms enclosed by the brackets in
  (\ref{up1}).   Since $\g$ breaks $E_8$ down to
  ${\cal G}_\g$, it follows that $\g$ acts trivially on those
  $E_8$ root vectors corresponding to the bracketed representations,
  and non-trivially otherwise.
  We notice that the ultimate representations in
  (\ref{up1}) and (\ref{up2}) are the same.  Thus,
  ${\cal H}=SU_6\times SU_3\times U_1$ satisfies the first
  necessary condition for the indicated branchings to be
  consistently entwined. We will analyze the second necessary
  condition, the absence of non-curable four-dimensional
  anomalies, shortly.

 \begin{figure}
 \begin{center}
 \includegraphics[width=3in,angle=0]{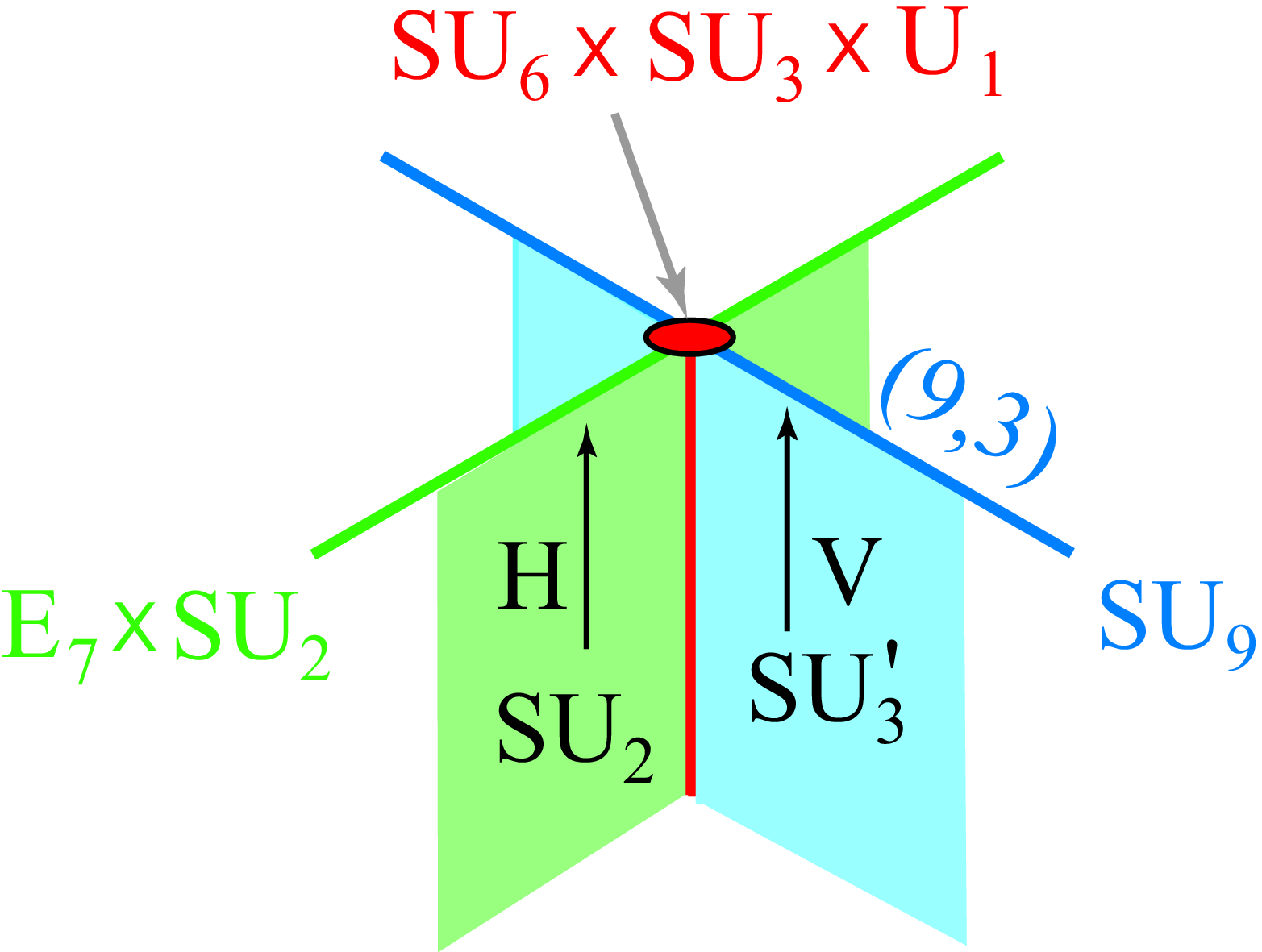} \\[.2in]
 \caption{The neighborhood of one of the $\a\b\g$-invariant
 four-plane intersections inside the upper end-of-the-world.
 Indicated in this diagram are the $E_8$ subgroups ${\cal G}_\a=E_7\times SU_2$
 and ${\cal G}_\g=SU_9$ which survive on the $\a\b$-invariant
 six-plane (the green line) and the $\b\g$-invariant six-plane
 (the blue line), respectively.  Also indicated is the
 representation of the six-dimensional twisted hypermultiplet
 needed to cure six-dimensional anomalies locally on the
 $\b\g$-plane.}
 \label{upcorner}
 \end{center}
 \end{figure}

 In generic situations, one way to analyze the problem of finding entwined branchings
 is to scan the lists of
 subgroups, depth by depth by referring to the tables in \cite{patera} or \cite{slansky},
 and look for ostensible matches.  One then has to carefully
 evaluate the branching rules in order to see if
 the selected higher-depth common group is, in fact properly entwined.
 One illustrative example is the following.  If we were seeking
 an entwined branching which involved as intermediaries
 the $E_8$ subgroups  $(\,SO_{16}\,,\,SU_9\,)$, we would discover
 that each of these group have $SU_8\times U_1$ subgroups.
 Thus, we would consider the possibility
 $(\,SO_{16}\,,\,SU_9\,|\,SU_8\times U_1\,)$.
 In this case, however, the ultimate representations do not coincide,
 as can be verified by direct computation.
 We conclude, therefore, that the group $SU_8\times U_1$ cannot be properly entwined
 inside of $E_8$ by the subgroups $SO_{16}$ and $SU_9$.

 Two convenient ways of exhibiting some of the relevant branching information
 described by (\ref{up1}) and (\ref{up2}) is to use branching
 diagrams or branching tables, two tools which were introduced in
 \cite{4d1}.
 For the case at hand, the relevant branching diagram and
 branching table are shown in Table \ref{upbranch}.
 \begin{table}
 \begin{center}
 \includegraphics[width=3in,angle=0]{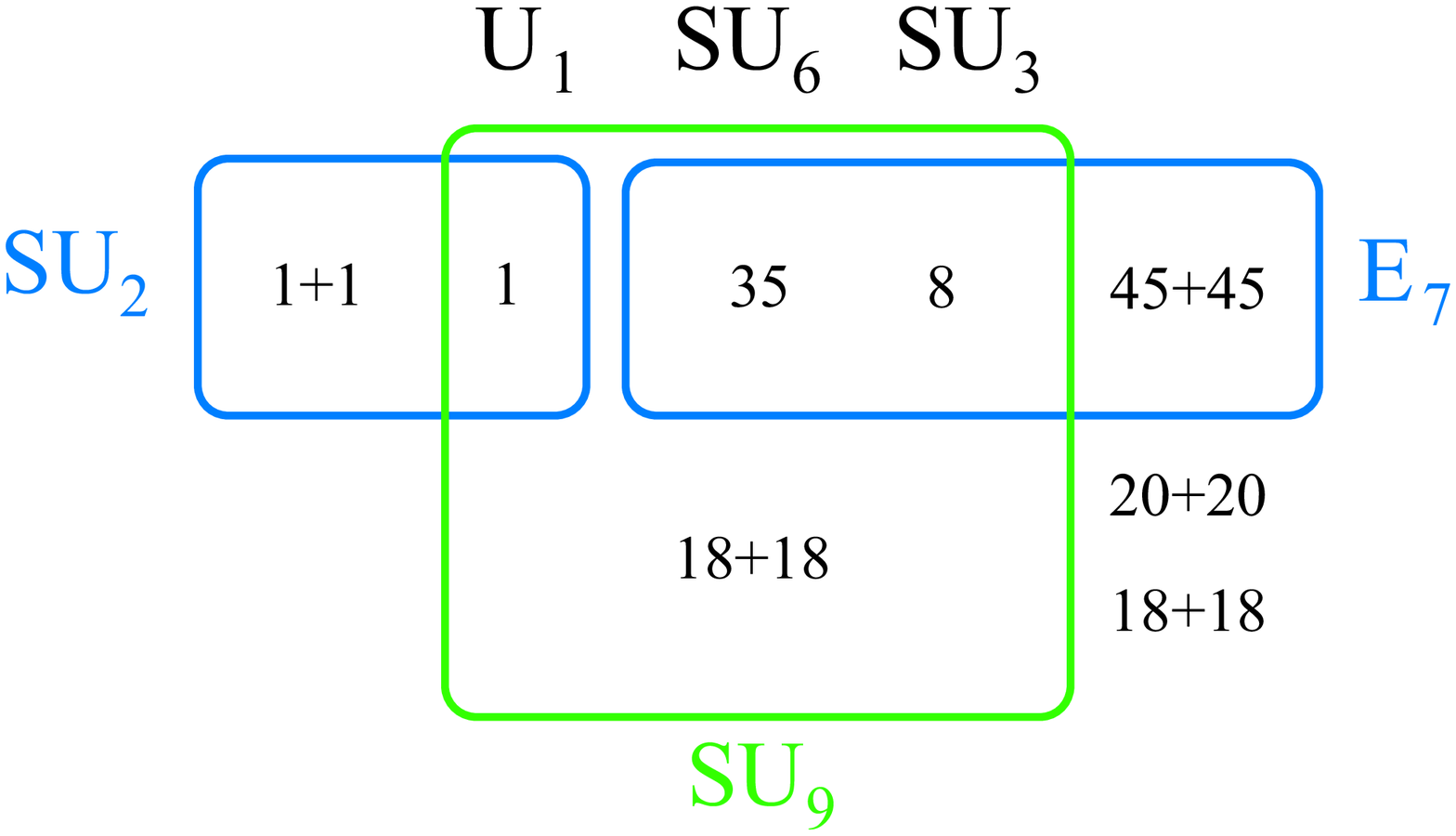}\\[.5in]
 \begin{tabular}{|c||c|c|c|}
 \hline
 & \hspace{.5in} & \hspace{.5in} & \hspace{.5in} \\[-.1in]
 {\bf 248} & $\a$ & $\b$ & $\g$ \\[.1in]
 \hline
 &&&\\[-.1in]
 $(\,{\bf 35}\,,\,{\bf 1}\,)_0$ & + & + & + \\[.1in]
 $(\,{\bf 1}\,,\,{\bf 8}\,)_0$ & + & + & + \\[.1in]
 $(\,{\bf 1}\,,\,{\bf 1}\,)_0$ & $+$ & + & + \\[.1in]
 $(\,{\bf 1}\,,\,{\bf 1}\,)_{-6}\,\oplus
 (\,{\bf 1}\,,\,{\bf 1}\,)_{+6}$ & $+$ & + & 1/3 \\[.1in]
 $(\,{\bf 15}\,,\,{\bf \bar{3}}\,)_0\,\oplus
 (\,{\bf \bar{15}}\,,\,{\bf 3}\,)_0$ & + & + & 1/3 \\[.1in]
 $(\,{\bf 20}\,,\,{\bf 1}\,)_{+3}\,\oplus
 (\,{\bf 20}\,,\,{\bf 1}\,)_{-3}$ & $-$ & + & 1/3 \\[.1in]
 $(\,{\bf 6}\,,\,{\bf 3}\,)_{-3}\,\oplus
 (\,{\bf \bar{6}}\,,\,{\bf \bar{3}}\,)_{+3}$ & $-$ & + & 1/3 \\[.1in]
 $(\,{\bf 6}\,,\,{\bf 3}\,)_{+3}\,\oplus
 (\,{\bf \bar{6}}\,,\,{\bf \bar{3}}\,)_{-3}$ & $-$ & + & + \\[.1in]
 \hline
 &&&\\[-.1in]
 $E_8\to SU_6\times SU_3\times U_1$ & $E_7\times SU_2$ & $E_8$ & $SU_9$ \\[.1in]
 \hline
 \end{tabular}\\[.5in]
 \caption{Embedding diagram and branching table describing
 $E_8\to SU_6\times SU_3\times U_1$.}
 \label{upbranch}
 \end{center}
 \end{table}
 The branching diagram is a simplified ``map" of the group $E_8$,
 showing, by dimensionality of subspaces, how the various
 subgroups are embedded.  The numbers in this diagram each refer to the
 dimensionality of one of the representations of ${\cal H}$
 included in the representation sums in (\ref{up1}) or,
 equivalently, (\ref{up2}).  The branching table indicates the
 action of the generating elements $\a$, $\b$ and $\g$ on the
 $E_8$ root lattice, by showing how these elements act on the
 representation indices associated with fields taking
 values in those representations of ${\cal H}$.

 From the information included in the branching table and the
 embedding diagram, it is straightforward to determine the
 spectrum which arises from the ten-dimensional $E_8$ fields
 which survive projection at the four-dimensional intersection.
 This is done by decomposing the ten-dimensional vector fields
 into four dimensional vector and scalar fields, and then
 considering the combination of the tensorial action induced by
 the quotient group elements, via their action on the spacetime
 coordinates, with the additional action on the representation
 indices associated with these same fields as induced by the
 action of the quotient group on the $E_8$ lattice.  The upshot, for
 the ten-dimensional fields, is that representations in
 the branching table transforming
 under $(\a,\b,\g)$ as $(+,+,+)$ supply $N=1$ vector
 multiplets.  In the next case, representations transforming as $(-,+,+)$
 supply $N=1$ chiral multiplets transforming according to the
 indicated representation.
 The remaining cases, $(+,+,\pm 1/3)$ and $(-,+,\mp 1/3)$ correspond to
 complex representations of the sort
 ${\bf{\cal R}}\oplus{\bf \bar{\cal R}}$.
 These also supply chiral multiplets, but the representation is
 truncated to ${\bf {\cal R}}$ or to ${\bf \bar{\cal R}}$, depending
 on the respective sign on the $1/3$ which appears in the branching table.

 It is now straightforward to read off of the branching table in
 Figure \ref{upbranch} the contribution to the four dimensional
 intersection spectrum which arises from the $E_8$ fields.  For
 the case at hand, using the rules described in the previous
 paragraph, we determine the spectrum indicated in the first
 column of Table \ref{uppers}.  The rational prefactors which
 appear in that table are distribution coefficients which we
 need to include in the computation of the four-dimensional anomaly.
 These are described below.  Notice that there are other
 contributions to the four-dimensional spectrum apparent
 in Figure \ref{uppers}.  Notably, we have six-dimensional twisted
 fields which need to be considered.  We describe these fields
 presently.

 As described above, in order to cancel the six-dimensional anomaly on the
 $\b\g$-invariant six-planes, we must include $SU_3'$ Yang-Mills supermultiplets
 on the intersecting $\g$-invariant seven-plane.  We must also
 include six-dimensional ``twisted" hypermultiplets on the $\b\g$-planes
 themselves, transforming as $(\,{\bf 9}\,,\,{\bf 3}\,)$ under
 $SU_9\times SU_3'$.  These reduce at the four-dimensional
 intersection into one $N=1$ chiral and one $N=1$ anti-chiral multiplet
 transforming as determined by
 the following branching,
 \brr SU_9\times SU_3' &\longrightarrow&
      SU_6\times SU_3\times SU_3'\times U_1
      \nonumber\\[.1in]
      (\,{\bf 9}\,,\,{\bf 3}\,) &\longrightarrow&
      (\,{\bf 6}\,,\,{\bf 1}\,,\,{\bf 3}\,)_{-1}\,\oplus\,
      (\,{\bf 1}\,,\,{\bf 3}\,,\,{\bf 3}\,)_{+2} \,.
 \label{z3twist}\err
 On the $\a\b\g$-invariant four-planes, however, the $\Z_2$ generator
 $\a$ projects this to one chiral multiplet (i.e. we project out the anti-chiral
 multiplet).  This explains the fields which appear in the
 ``6D" column in Figure \ref{uppers}.
 The collective situation at one of the upstairs four-planes
 is illustrated in Figure \ref{upcorner}.  In that figure
 we can see the variety of fixed-planes which mutually intersect
 at the given four-plane.  We can also see the effective gauge
 group and the spectrum of twisted fields associated with the
 each of these planes.

 \begin{table}
 \begin{center}
 \begin{tabular}{|c||c|c|c|c|}
 \hline
 &
 \hspace{1in} &
 \hspace{1in} &
 \hspace{1in} &
 \hspace{1in} \\[-.1in]
 & 10D & 7D & 6D & 4D \\[.1in]
 \hline
 &&&&\\[-.1in]
 Chiral & $\ft{1}{12}\,(\,{\bf 1}\,,\,{\bf 1}\,,\,{\bf 1}\,)_{-6}$ &
 $\ft{1}{6}\,(\,{\bf 1}\,,\,{\bf 1}\,,\,{\bf 1}\,)_0$ &
 $\ft14\,(\,{\bf 6}\,,\,{\bf 1}\,,\,{\bf 3}\,)_{-1}$ & \\[.1in]
 & $\ft{1}{12}\,(\,{\bf 15}\,,\,{\bf \bar{3}}\,,\,{\bf 1}\,)_{0}$ &
 $\ft{1}{6}\,(\,{\bf 1}\,,\,{\bf 1}\,,\,{\bf 1}\,)_{+3}$ &
 $\ft14\,(\,{\bf 1}\,,\,{\bf 3}\,,\,{\bf 3}\,)_{+2}$ & \\[.1in]
 & $\ft{1}{12}\,(\,{\bf 20}\,,\,{\bf 1}\,,\,{\bf 1}\,)_{+3}$ &
 $\ft{1}{6}\,(\,{\bf 1}\,,\,{\bf 1}\,,\,{\bf 1}\,)_{-3}$ & & \\[.1in]
 & $\ft{1}{12}\,(\,{\bf 6}\,,\,{\bf 3}\,,\,{\bf 1}\,)_{-3}$ & & & \\[.1in]
 & $\ft{1}{12}\,(\,{\bf 6}\,,\,{\bf 3}\,,\,{\bf 1}\,)_{+3}$ & & & \\[.1in]
 & $\ft{1}{12}\,(\,{\bf \bar{6}}\,,\,{\bf \bar{3}}\,,\,{\bf 1}\,)_{-3}$ & & &
 \\[.1in]
 \hline
 &&&& \\[-.1in]
 Vector & $\ft{1}{12}\,(\,{\bf 35}\,,\,{\bf 1}\,,\,{\bf 1}\,)_0$ &
 $\ft{1}{8}\,(\,{\bf 1}\,,\,{\bf 1}\,,\,{\bf 8}\,)_0$ & & \\[.1in]
 & $\ft{1}{12}\,(\,{\bf 1}\,,\,{\bf 8}\,,\,{\bf 1}\,)_0$ & & & \\[.1in]
 & $\ft{1}{12}\,(\,{\bf 1}\,,\,{\bf 1}\,,\,{\bf 1}\,)_0$ & & & \\[.1in]
 \hline
 \end{tabular} \\[.2in]
 \caption{The effective spectrum, in terms of $N=1$ superfields, as seen by one of the
 upstairs $\a\b\g$ four-planes, in terms of representations of
 $SU_6\times SU_3\times SU_3'\times U_1$.  The rational numbers which appear in this
 table are the distribution coefficients needed to amend the index theory
 computation of the local four-dimensional anomalies.}
 \label{uppers}
 \end{center}
 \end{table}
 The seven-dimensional twisted fields, localized on the
 $\a$-invariant and $\g$-invariant seven-planes also contribute effectively to the
 local four-dimensional spectrum.  However, these fields do not
 contribute chirally, and are not relevant to the four-dimensional
 anomaly discussion.

 A four dimensional anomaly arises because the higher-dimensional
 fermion fields couple in a chiral fashion locally, at the four-dimensional
 intersection, to the gauge currents associated with ${\cal H}$.
 As is well-known, the index-theory
 computation of the relevant anomalies needs to be modified by the
 incorporation of appropriate distribution divisors.
 For instance, since a given $\b\g$-invariant six-plane plane shares four
 $\a\b\g$-invariant four-planes as subspaces, the four-dimensional anomaly associated
 with a given $\b\g$ plane includes a distribution divisor
 of 4.  Similarly, the ten-dimensional contribution to the
 four-dimensional anomaly should include a distribution divisor
 of twelve.  This is obtained from the observation that each
 $\b$-invariant ten-plane includes twelve indistinguishable
 $\a\b\g$-invariant four planes, as is evident in Figure
 \ref{endtori}.

 The charged
 \footnote{By charged we mean terms which have a nonvanishing
 $U(1)$ charge.}
 chiral spectrum ``seen" by a given $\a\b\g$ four-plane,
 in terms of $SU_6\times SU_3\times SU_3'\times U_1$
 representations, consists of the following terms derived from ten dimensions,
 \brr && \frac{1}{12}\,\bpl\,
      (\,{\bf 1}\,,\,{\bf 1}\,,\,{\bf 1}\,)_{-6}\,\oplus\,
      (\,{\bf 20}\,,\,{\bf 1}\,,\,{\bf 1}\,)_{+3}\,\oplus\,
      (\,{\bf 6}\,,\,{\bf 3}\,,\,{\bf 1}\,)_{-3}\,\bpr
 \label{chiralten}\err
 and also the following terms derived from six dimensions,
 \brr  \frac{1}{4}\,\bpl\,
      (\,{\bf 6}\,,\,{\bf 1}\,,\,{\bf 3}\,)_{-1}\,\oplus\,
      (\,{\bf 1}\,,\,{\bf 3}\,,\,{\bf 3}\,)_{+2}\,\bpr \,.
 \label{chiralsix}\err
 In each case, the rational pre-factor is the anomaly distribution
 coefficient.
 Using the chiral spectrum shown in (\ref{chiralten})
 and (\ref{chiralsix}), we can now compute the
 the four-dimensional gauge anomaly seen by one of the $\a\b\g$
 planes.  The precise technology for doing this is explained in
 Appendix \ref{anom4d}.  The relevant anomalies are the
 gauge anomalies for the simple factors $SU_6$, $SU_3$ and
 $SU_3'$, the gauge anomaly for the $U_1$ factor, and the mixed
 anonaly involving the $U_1$ factor.  These can be computed using
 (\ref{ganom}) and (\ref{manom}).  We find,
 \brr I(\,SU_6\,) &=&
      \ft{1}{12}\,\bpl\, 1\,(+3)\,I_2({\bf 20})+3\,(-3)\,I_2({\bf 6})\,\bpr
      +\ft14\,\bpl\,3(-1)\,I_2({\bf 6})\,\bpr
      \nonumber\\[.1in]
      I(\,SU_3\,) &=& \ft{1}{12}\,\bpl\,6\,(-3)\,I_2(\,{\bf 3}\,)\,\bpr
      +\ft14\,\bpl\,3\,(+2)\,I_2(\,{\bf 3}\,)\,\bpr
      \nonumber\\[.1in]
      I(\,SU_3'\,) &=&
      \ft14\,\bpl\,6\,(-1)+3\,(+2)\,\bpr\,I_2(\,{\bf 3}\,)
      \nonumber\\[.1in]
      I(\,U_1\,)_{\rm GAUGE} &=& \ft{1}{12}\,\bpl\,
      1\,(-6)^3+20\,(+3)^3+18\,(-3)^3\,\bpr
      +\ft14\,\bpl\,
      18\,(-1)^3+9\,(+2)^3\,\bpr
      \nonumber\\[.1in]
      I(U_1)_{\rm MIXED} &=& \ft{1}{12}\,\bpl\,
      1\,(-6)+20\,(+3)+18\,(-3)\,\bpr
      +\ft14\,\bpl\,
      18\,(-1)+9\,(+2)\,\bpr
      \,,
 \label{anoms}\err
 where $I_2({\bf{\cal R}})$ denotes the second index associated
 with the representation ${\bf{\cal R}}$.
 For $SU({\bf N})$, the second index of the fundamental ${\bf N}$ representations are always
 unity, i.e. $I_2({\bf N})=1$.
 Thus, for $SU_6$, we have $I_2(\,{\bf 6}\,)=1$, and
 for each of the two $SU(3)$ factors we have $I_2(\,{\bf 3}\,)=1$.
 The ${\bf 20}$ is the three-index antisymmetric tensor
 representation of $SU_6$.
 Therefore, we use the algorithm explained in Appendix \ref{indices} to compute
 $I_2(\,{\bf 20}\,)=6$.  Using these results,  it is easy to show that
 each of the five anomaly expressions in (\ref{anoms}) vanishes
 identically.  Thus, we have satisfied the second non-trivial check that
 the entwined branching
 $(\,E_7\times SU_2\,,\,SU_9\,|\,SU_6\times SU_3\times U_1\,)$ is, in fact
 consistent.

 \subsection{Downstairs Vertices}
 \begin{figure}
 \begin{center}
 \includegraphics[width=3.2in,angle=0]{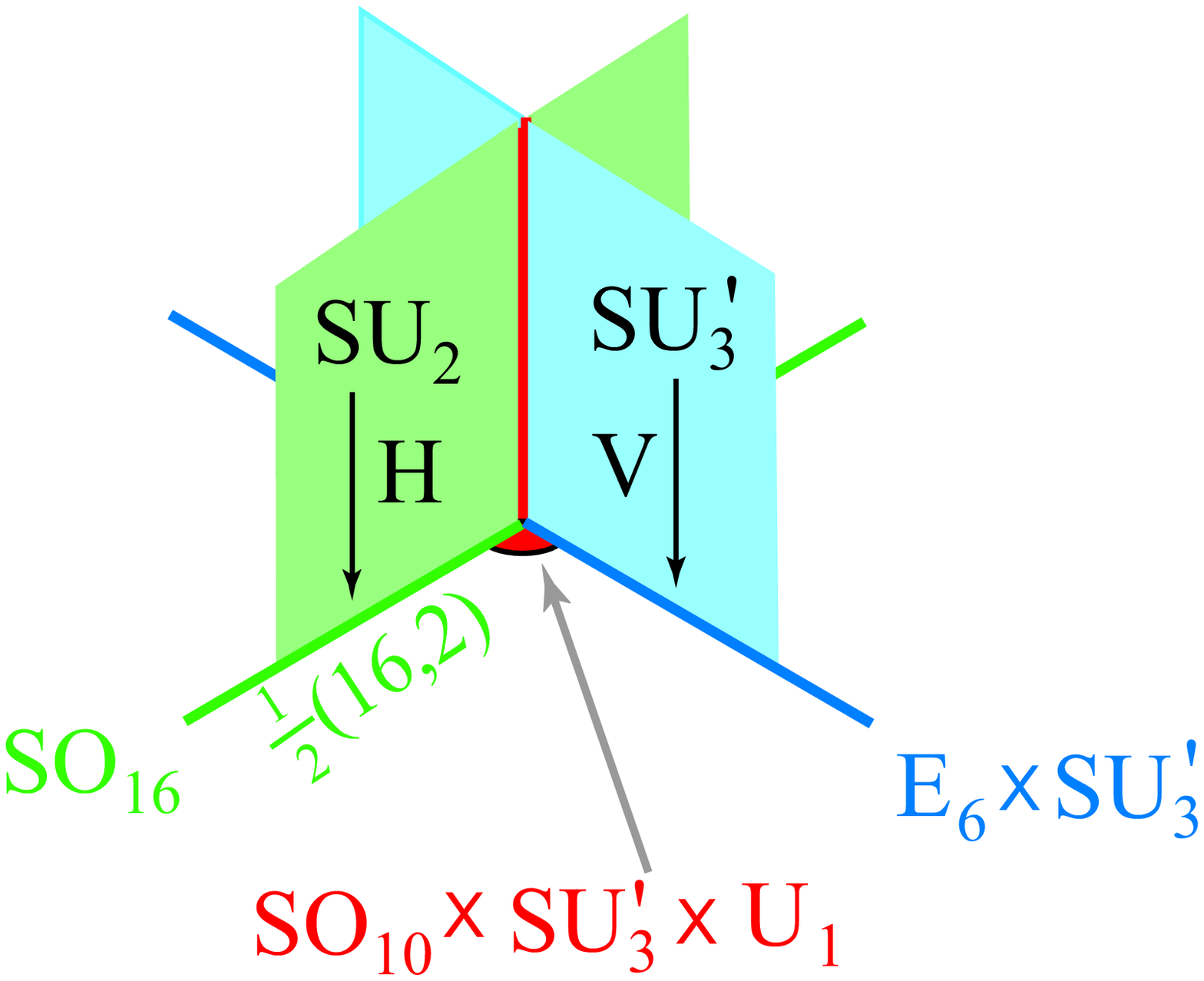} \\[.2in]
 \caption{The neighborhood of one of the $\a\b\g$-invariant
 four-plane intersections inside the lower end-of-the-world.
 Indicated in this diagram are the $E_8$ subgroups ${\cal G}_\a=SO_{16}$
 and ${\cal G}_\g=E_6\times SU_3$ which survive on the $\a\b$-invariant
 six-plane (the green line) and the $\b\g$-invariant six-plane
 (the blue line), respectively.  Also indicated is the
 representation of the six-dimensional twisted hypermultiplet
 needed to cure six-dimensional anomalies locally on the
 $\b\g$-plane.}
 \label{downcorner}
 \end{center}
 \end{figure}
 Owing to global $G$-flux conservation, we are obligated to incorporate
 on the downstairs vertices consistently entwined branchings
 complimentary to that discussed above.  We have determined that
 the following is satisfactory
 $(\,SO_{16}\,,\,E_6\times SU_3'\,|\, SO_{10}\times SU_3'\times U_1'\,)$.
 We suspect this is the unique solution.
 In this subsection we will analyze the branching and anomaly
 questions pertaining to this choice in a manner
 analogous to the discussion in the previous subsection.
 Since the reasoning is identical, we will be comparatively brief.
 For the case at hand, the $(\,\a\,,\,\g\,)$ branching is given by
 \footnote{Note that $SO_{10}\times SU_3'\times U_1'$ is at depth-two inside
 $SO_{16}$; branching through successive maximal subgroups, we have
  $E_8\to SO_{16}\to (\,SO_{10}\times SU_4\,)\to SO_{10}\times SU_3'\times
  U_1'$.  The step involving $SO_{10}\times SU_4$ representations
  is suppressed in (\ref{bran1}).}
 \brr E_8 &\stackrel{\a}{\longrightarrow}& SO_{16}
      \nonumber\\[.1in]
      &\stackrel{\g}{\longrightarrow}& SO_{10}\times SU_3'\times
      U_1'
      \nonumber\\[.2in]
      {\bf 248} &\stackrel{\a}{\longrightarrow}&
      [\,{\bf 120}\,]\,\oplus\,{\bf 128}
      \nonumber\\[.1in]
      &\stackrel{\g}{\longrightarrow}&
      [\,(\,{\bf 45}\,,\,{\bf 1}\,)_0\,\oplus\,
      (\,{\bf 1}\,,\,{\bf 8}\,)_0\,\oplus\,
      (\,{\bf 1}\,,\,{\bf 1}\,)_0
      \nonumber\\[.1in]
      & & \oplus\,(\,{\bf 10}\,,\,{\bf 3}\,)_{-2}\,\oplus\,
      (\,{\bf 10}\,,\,{\bf \bar{3}}\,)_{+2}\,\oplus\,
      (\,{\bf 1}\,,\,{\bf 3}\,)_{+4}\,\oplus\,
      (\,{\bf 1}\,,\,{\bf \bar{3}}\,)_{-4}\,]
      \nonumber\\[.1in]
      & & \,\oplus\,(\,{\bf 16}\,,\,{\bf 1}\,)_{-3}\,\oplus\,
      (\,{\bf 16}\,,\,{\bf 3}\,)_{+1}\,\oplus\,
      (\,{\bf \bar{16}}\,,\,{\bf 1}\,)_{+3}\,\oplus\,
      (\,{\bf \bar{16}}\,,\,{\bf \bar{3}}\,)_{-1} \,.
  \label{bran1}\err
  Next, for the $(\g,\a)$ branching, we find
  \brr E_8 &\stackrel{\g}{\longrightarrow}& E_6\times SU_3'
      \nonumber\\[.1in]
      &\stackrel{\a}{\longrightarrow}& SO_{10}\times SU_3'\times
      U_1'
      \nonumber\\[.2in]
      {\bf 248} &\stackrel{\g}{\longrightarrow}&
      [\,(\,{\bf 78}\,,\,{\bf 1}\,)\,\oplus\,
      (\,{\bf 1}\,,\,{\bf 8}\,)\,]\,\oplus\,
      (\,{\bf 27}\,,\,{\bf 3}\,)\,\oplus\,
      (\,{\bf \bar{27}}\,,\,{\bf \bar{3}}\,)
      \nonumber\\[.2in]
      &\stackrel{\a}{\longrightarrow}&
      [\,(\,{\bf 45}\,,\,{\bf 1}\,)_0\,\oplus\,
      (\,{\bf 16}\,,\,{\bf 1}\,)_{-3}\,\oplus\,
      (\,{\bf \bar{16}}\,,\,{\bf 1}\,)_{+3}\,\oplus\,
      (\,{\bf 1}\,,\,{\bf 1}\,)_0\,\oplus\,
      (\,{\bf 1}\,,\,{\bf 8}\,)_0\,]
      \nonumber\\[.1in]
      & & \oplus\,(\,{\bf 16}\,,\,{\bf 3}\,)_{+1}\,\oplus\,
      (\,{\bf 10}\,,\,{\bf 3}\,)_{-2}\,\oplus\,
      (\,{\bf 1}\,,\,{\bf 3}\,)_{+4}
      \nonumber\\[.1in]
      & & \oplus\,(\,{\bf \bar{16}}\,,\,{\bf \bar{3}}\,)_{-1}\,\oplus\,
      (\,{\bf 10}\,,\,{\bf \bar{3}}\,)_{+2}\,\oplus\,
      (\,{\bf 1}\,,\,{\bf \bar{3}}\,)_{-4}
 \label{bran2}\err
 Once again, notice that the ultimate representations in
 (\ref{bran1}) and (\ref{bran2}) coincide.  As described above,
 this is a necessary condition on entwined branchings.
 Relevant aspects of this branching are usefully exhibited
 in the embedding diagram and branching table shown in
 Figure \ref{velky}.
 \begin{table}
 \begin{center}
 \includegraphics[width=3in,angle=0]{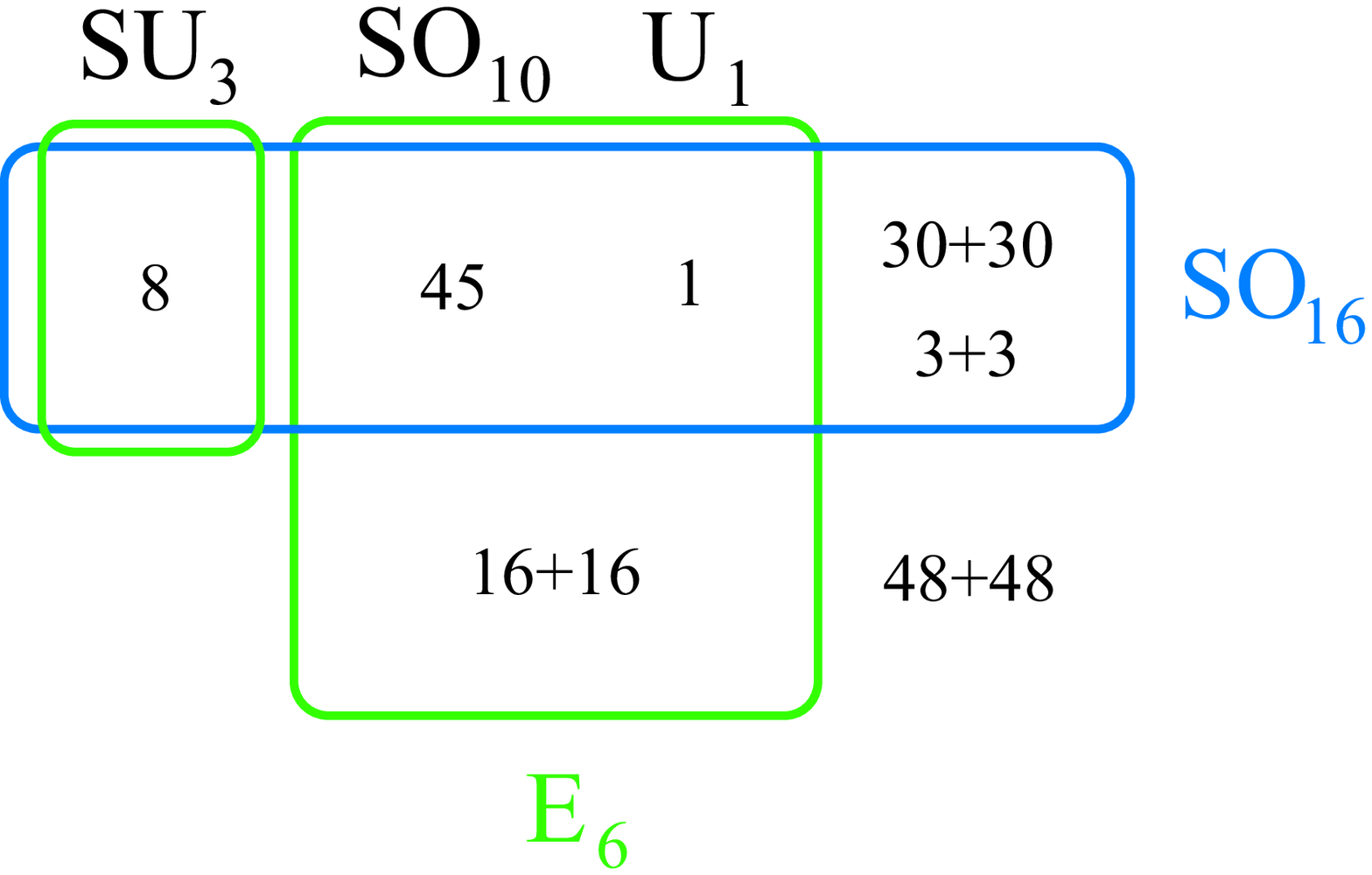}\\[.5in]
 \begin{tabular}{|c||c|c|c|}
 \hline
 & \hspace{.5in} & \hspace{.5in} & \hspace{.5in} \\[-.1in]
 {\bf 248} & $\a$ & $\b$ & $\g$ \\[.1in]
 \hline
 &&&\\[-.1in]
 $(\,{\bf 45}\,,\,{\bf 1}\,)_0$ & + & + & + \\[.1in]
 $(\,{\bf 1}\,,\,{\bf 8}\,)_0$ & + & + & + \\[.1in]
 $(\,{\bf 1}\,,\,{\bf 1}\,)_0$ & + & + & + \\[.1in]
 $(\,{\bf 10}\,,\,{\bf 3}\,)_{-2}\,\oplus\,
 (\,{\bf \bar{10}}\,,\,{\bf \bar{3}}\,)_{+2}$ & + & + & 1/3 \\[.1in]
 $(\,{\bf 1}\,,\,{\bf 3}\,)_{+4}\,\oplus\,
 (\,{\bf \bar{1}}\,,\,{\bf \bar{3}}\,)_{-4}$ & + & + & 1/3 \\[.1in]
 $(\,{\bf 16}\,,\,{\bf 1}\,)_{-3}\,\oplus\,
 (\,{\bf \bar{16}}\,,\,{\bf 1}\,)_{+3}$ & $-$ & + & + \\[.1in]
 $(\,{\bf 16}\,,\,{\bf 3}\,)_{+1}\,\oplus\,
 (\,{\bf \bar{16}}\,,\,{\bf \bar{3}}\,)_{-1}$ & $-$ & + & 1/3 \\[.1in]
 \hline
 &&&\\[-.1in]
 $E_8\to SO_{10}\times SU_3'\times U_1$ & $SO_{16}$ & $E_8$ & $E_6\times SU_3'$ \\[.1in]
 \hline
 \end{tabular}\\[.5in]
 \caption{Embedding diagram and branching table describing
 $E_8\to SO_{10}\times SU_3'\times U_1$.}
 \label{velky}
 \end{center}
 \end{table}

 In order to cancel local anomalies on the
 $\a\g$-invariant six-planes, we must include $SU_2$ Yang-Mills supermultiplets
 on the intersecting $\a$-invariant seven-plane.  We must also
 include twisted hypermultiplets on the $\a\g$-planes
  themselves, transforming as $\ft12(\,{\bf 16}\,,\,{\bf 2}\,)$ under
 $SO_{16}\times SU_2$.  These reduce at the four-dimensional
 intersection into one $N=1$ chiral multiplet
 \footnote{The factor of $\ft12$ on the hypermultiplet
 representation serves to remove in the decomposition the
 antichiral multiplets, which, owing to the pseudoreality of the
 representation, is equivalent, via charge conjugation, to
 a second set of chiral multiplets.}
 transforming under $SO_{10}\times SU_3'\times U_1'$ as determined by
 the following branching,
 \brr SO_{16}\times SU_2 &\longrightarrow&
      SO_{10}\times SU_4\times SU_2
      \nonumber\\[.1in]
      &\longrightarrow&
      SO_{10}\times SU_3'\times SU_2\times U_1'
      \nonumber\\[.2in]
      (\,{\bf 16}\,,\,{\bf 2}\,) &\longrightarrow&
      (\,{\bf 10}\,,\,{\bf 1}\,,\,{\bf 2}\,)\,\oplus\,
      (\,{\bf 1}\,,\,{\bf 6}\,,\,{\bf 2}\,)
      \nonumber\\[.1in]
      &\longrightarrow&
      (\,{\bf 10}\,,\,{\bf 1}\,,\,{\bf 2}\,)_0\,\oplus\,
      (\,{\bf 1}\,,\,{\bf 3}\,,\,{\bf 2}\,)_{+2}\,\oplus\,
      (\,{\bf 1}\,,\,{\bf \bar{3}}\,,\,{\bf 2}\,)_{-2}
 \label{z2twist}\err
 On the $\a\b\g$-invariant four-planes, the $\Z_3$ generator
 $\g$ acts trivially on these six-dimensional twisted fields and,
 does not serve to reduce further the degrees of freedom.
 Since a given $\a\b$-invariant six-plane shares three
 $\a\b\g$-invariant four-planes as subspaces, the four-dimensional anomaly associated
 with fields on a given $\b\g$-invariant six-plane includes a distribution divisor
 of three. Therefore, these fields contribute to the effective
 four-dimensional spectrum those fields indicated in the ``6D" column of
 Table \ref{lowers}.
 The collective situation at one of the downstairs four-planes
 is illustrated in Figure \ref{downcorner}.

 \begin{table}
 \begin{center}
 \begin{tabular}{|c||c|c|c|c|}
 \hline
 &
 \hspace{1in} &
 \hspace{1in} &
 \hspace{1in} &
 \hspace{1in} \\[-.1in]
 & 10D & 7D & 6D & 4D \\[.1in]
 \hline
 &&&&\\[-.1in]
 Chiral &
 $\ft{1}{12}\,(\,{\bf 10}\,,\,{\bf 3}\,,\,{\bf 1}\,)_{-2}$ &
 $\ft{1}{8}\,(\,{\bf 1}\,,\,{\bf 8}\,,\,{\bf 1}\,)_0$ &
 $\ft13\,(\,{\bf 10}\,,\,{\bf 1}\,,\,{\bf 2}\,)_0$ & \\[.1in]
 & $\ft{1}{12}\,(\,{\bf 1}\,,\,{\bf 3}\,,\,{\bf 1}\,)_{+4}$ & &
 $\ft13\,(\,{\bf 1}\,,\,{\bf 3}\,,\,{\bf 2}\,)_{+2}$ & \\[.1in]
 & $\ft{1}{12}\,(\,{\bf 16}\,,\,{\bf 3}\,,\,{\bf 1}\,)_{+1}$ & &
 $\ft{1}{3}\,(\,{\bf 1}\,,\,{\bf \bar{3}}\,,\,{\bf 2}\,)_{-2}$ & \\[.1in]
 & $\ft{1}{12}\,(\,{\bf 16}\,,\,{\bf 1}\,,\,{\bf 1}\,)_{-3}$ & & & \\[.1in]
 & $\ft{1}{12}\,(\,{\bf \bar{16}}\,,\,{\bf 1}\,,\,{\bf 1}\,)_{+3}$ & & & \\[.1in]
 \hline
 &&&& \\[-.1in]
 Vector & $\ft{1}{12}\,(\,{\bf 45}\,,\,{\bf 1}\,,\,{\bf 1}\,)_0$ &
 $\ft{1}{6}\,(\,{\bf 1}\,,\,{\bf 1}\,,\,{\bf 3}\,)_0$ & & \\[.1in]
 & $\ft{1}{12}\,(\,{\bf 1}\,,\,{\bf 8}\,,\,{\bf 1}\,)_0$ & & & \\[.1in]
 & $\ft{1}{12}\,(\,{\bf 1}\,,\,{\bf 1}\,,\,{\bf 1}\,)_0$ & & & \\[.1in]
 \hline
 \end{tabular} \\[.2in]
 \caption{The effective spectrum, in terms of $N=1$ superfields, as seen by one of the
 downstairs $\a\b\g$ four-planes, in terms of representations of
 $SO_{10}\times SU_3'\times SU_2\times U_1'$.}
 \label{lowers}
 \end{center}
 \end{table}

 The charged chiral spectrum seen by a given $\a\b\g$-invariant four-plane
 consists of the following terms derived from ten dimensions,
 \brr && \frac{1}{12}\,\bpl\,
      (\,{\bf 10}\,,\,{\bf 3}\,,\,{\bf 1}\,)_{-2}\,\oplus\,
      (\,{\bf 1}\,,\,{\bf 3}\,,\,{\bf 1}\,)_{+4}\,\oplus\,
      (\,{\bf 16}\,,\,{\bf 3}\,,\,{\bf 1}\,)_{+1}\,\bpr \,.
 \label{below}\err
 The terms derived from six dimensions have no anomaly.
 In (\ref{below}) the pre-factor one-twelfth is the anomaly
 distribution coefficient.  As described above, this
 derives from the fact that there are twelve indistinguishable
 $\a\b\g$-invariant four-planes
 within each $\b$-invariant ten-plane.
 Using the chiral spectrum shown in (\ref{below}), we can now compute the
 the four-dimensional gauge anomaly seen by one of the $\a\b\g$
 planes.  This is done according to the rules explained in
 Appendix \ref{anom4d}.  The relevant anomalies are
 the gauge anomalies for the simple factors $SO_{10}$ and
 $SU_3'$, the gauge anomaly for the $U_1'$ factor and the mixed
 anomaly involving the $U_1'$ factor.
 \brr I(\,SO_{10}\,) &=&
      \ft{1}{12}\,\bpl\,
      3\,(-2)\,I_2({\bf 10})
      +3\,(+1)\,I_2({\bf 16})\,\bpr
      \nonumber\\[.1in]
      I(\,SU_3'\,) &=& \ft{1}{12}\,\bpl\,
      10\,(-2)+1\,(+4)+16\,(+1)\,\bpr\,I_2(\,{\bf 3}\,)
      \nonumber\\[.1in]
      I(\,U_1\,)_{\rm GAUGE} &=& \ft{1}{12}\,\bpl\,
      30\,(-2)^3
      +3\,(+4)^3
      +48\,(+1)^3\,\bpr
      \nonumber\\[.1in]
      I(\,U_1\,)_{\rm MIXED} &=& \ft{1}{12}\,\bpl\,
      30\,(-2)
      +3\,(+4)
      +48\,(+1)\,\bpr \,.
 \label{anomsdown}\err
 The second indices for the fundamental representations
 {\bf 10} of $SO(10)$ and ${\bf 3}$ of $SU(3)$ are unity
 by definition.  Thus, $I_2({\bf 10})=1$ and $I_2({\bf 3})=1$
 in (\ref{anomsdown}). For integer $l$, the groups $SO(2l)$
 have elementary spinor representations
 with dimension $2^{l-1}$.
 These representations have second index $2^{l-4}$.  Thus,
 for $SO_{10}$, we have $I_2({\bf 16})=2$.
 Using these results,  we easily show that
 each of the four anomaly expressions in (\ref{anoms}) vanishes
 identically. Thus, we have satisfied the non-trivial check that
 the  entwined branching $(\,SO_{16}\,,\,E_6\times
 SU_3'\,|\,SO_{10}\times SU_3'\times U_1'\,)$ is, in fact,
 consistent.

 At this point we have verified that all local anomalies in this
 orbifold, including those concentrated on ten-dimensional
 fixed planes, six-dimensional fixed planes and also
 four-dimensional fixed planes have been eliminated.
 The four-dimensional anomalies have been analyzed at the
 24 $\a\b\g$-invariant fixed planes, twelve on the upper
 end-of-the-world, and twelve on the lower end-of-the-world.
 One might wonder about possible four-dimensional anomalies
 localized at the other two classes of four-dimensional intersections
 which occur in this orbifold.  One of these classes comprises the sixteen
 triple intersections of the $\a\g$-invariant six-planes
 (the grey spots in Figure \ref{endtori}).
 The other class comprises the twelve double intersections of the
 $\b\g$-invariant six-planes (the yellow spots in Figure
 \ref{endtori}).  In each of these cases the effective
 four-dimensional spectrum seen by these intersections is
 non-chiral.  For this reason there are no four-dimensional
 anomaly constraints associated with these intersections.
 The fact that the effective spectrum at these points is
 non-chiral is related to the fact that these intersection
 points are not $\Gamma$-invariants, this in contrast to the
 $\a\b\g$-invariant four-planes.  (It is for a similar reason
 that the effective spectrum associated with the orbifold
 described in \cite{4d1} is non-chiral.)

 Notably, despite the chiral nature of the
 locally-observed spectrum, our solution does not require
 any extra four-dimensional twisted fields to
 remove the four-dimensional intersection anomalies.
 This was unexpected.  In more general orbifolds
 we do expect that such four-dimensional local matter will
 be necessary.  In fact, the circumstance in which the
 four-dimensional intersection anomalies would be non-trivially
 cured by the addition of new fields localized at intersections
 would be especially interesting.

 \section{When Worlds Collide}
 If we consider a limit where all compact dimensions except the
 interval direction $x^{11}$ are taken small, we obtain a picture
 of two four-dimensional ends-of-the-world, connected by a
 five-dimensional  bulk.  We shall refer to this as the ``spindle"
 limit, since this describes a situation where the seven compact dimensions
 degenerate to a spindle shape.  Chiral matter living on the
 the ``upper world" (at the top of the spindle) transforms under
 $SU_6\times SU_3\times SU_3'\times U_1$, whereas chiral matter
 living on the ``lower world" (at the bottom of the spindle) transforms under
 $SO_{10}\times SU_3'\times SU_2\times U_1'$.  The precise matter
 content corresponding to these two worlds
 is obtained from three different sources, corresponding to the
 three different classes of ``neighborhoods" shown in
 Figure \ref{wheels}.

 Primarily, there is the chiral spectrum
 associated with the $\a\b\g$-invariant intersections
 analyzed in the previous section.  The
 four-dimensional chiral spectrum obtained in the spindle limit
 is obtained for the upper world from Table \ref{uppers} and
 for the lower world from Table \ref{lowers}.
 In the spindle limit, the contributions from the
 ten-dimensional twisted fields and from the six-dimensional
 twisted fields appear as chiral multiplets transforming
 according to representations indicated in those tables.
 However, as the compact space $X^7$ coalesces to a spindle, the associated anomaly
 distribution coefficients add up to unity.  This, in fact, is
 what justified those coefficients in the first place.
 From the spindle point-of-view the seven-dimensional fields
 appearing in Tables \ref{uppers} and \ref{lowers}
 become five-dimensional (bulk) fields.

 Next, there is the non-chiral spectrum associated with the
 the six-planes which do not have $\Gamma$-invariant subspaces.
 For instance, out of the nine $\b\g$-invariant six-planes,
 three of these intersect
 $\a\b$-invariant six-planes, and six do not.  The associated
 six-dimensional twisted fields on the upper world are projected as in
 (\ref{z3twist}) in each case.  However, in those three cases involving
 $\Gamma$-invariant intersections, the contribution to the
 four-dimensional spectrum consists exclusively of chiral multiplets
 transforming as indicated.
 (In those cases the antichiral components are projected out.)
 In the six remaining cases, where there are no $\Gamma$-invariant
 subspaces, the six-dimensional twisted fields provide
 complementary sets of chiral and anti-chiral multiplets, each transforming
 according to the representation on the right-hand side of
 (\ref{z3twist}).  By charge-conjugation, however, this is equivalent
 to one set of chiral multiplets transforming in that
 same way, and another set of chiral multiplets transforming
 in the conjugate representation.  Thus, the
 six $\b\g$-invariant six-planes without $\Gamma$-invariant subspaces
 provide a non-chiral sector consisting of twelve sets
 of chiral multiplets, six of which
 transform under $SU_6\times SU_3\times SU_3'\times U_1$
 as ${\bf{\cal R}}\equiv(\,{\bf 6}\,,\,{\bf 1}\,,\,{\bf 3}\,)_{-1}\,\oplus\,
 (\,{\bf 1}\,,\,{\bf 3}\,,\,{\bf 3}\,)_{+2}$, and six more
 transforming as the conjugate of this
 representation.   Adding these
 contributions together, we have nine sets of chiral
 fields transforming as ${\bf{\cal R}}$, and only
 six transforming as ${\bf\bar{{\cal R}}}$.

 We also have six-dimensional twisted fields on the $\a\b$-planes in the lower world.
 Each of the sixteen $\a\b$-planes on the lower world supports
 fields transforming as shown in (\ref{z2twist}).  In this case
 there is no further projection imparted at the $\Gamma$-invariant
 intersections. Therefore, the six-dimensional twisted fields in the
 lower world contributes to the four-dimensional spectrum sixteen
 indistinguishable non-chiral families.

 Combining all of the above, we have determined two
 complimentary {\it M}-theory ``worlds".  The respective four-dimensional
 chiral spectra, determined along the lines described above are
 summarized in Tables \ref{upper} and \ref{lower}.  These two
 worlds are connected by a five-dimensional ``bulk", which
 consists of minimal five-dimensional supergravity coupled to
 an $SU_2\times SU_3'$ gauge sector.

 \begin{table}
 \begin{center}
 \begin{tabular}{|c|c|}
 \hline
 &\\[-.1in]
 $(\,{\bf 1}\,,\,{\bf 1}\,,\,{\bf 1}\,)_{-6}$ &
 $9\,(\,{\bf 6}\,,\,{\bf 1}\,,\,{\bf 3}\,)_{-1}$ \\[.1in]
 $(\,{\bf 15}\,,\,{\bf \bar{3}}\,,\,{\bf 1}\,)_{0}$ &
 $6\,(\,{\bf \bar{6}}\,,\,{\bf 1}\,,\,{\bf \bar{3}}\,)_{+1}$ \\[.1in]
 $(\,{\bf 20}\,,\,{\bf 1}\,,\,{\bf 1}\,)_{+3}$ &
 $9\,(\,{\bf 1}\,,\,{\bf 3}\,,\,{\bf 3}\,)_{+2}$ \\[.1in]
 $(\,{\bf 6}\,,\,{\bf 3}\,,\,{\bf 1}\,)_{-3}$ &
 $6\,(\,{\bf 1}\,,\,{\bf \bar{3}}\,,\,{\bf \bar{3}}\,)_{-2}$ \\[.1in]
 $(\,{\bf 6}\,,\,{\bf 3}\,,\,{\bf 1}\,)_{+3}$ &  \\[.1in]
 $(\,{\bf \bar{6}}\,,\,{\bf \bar{3}}\,,\,{\bf 1}\,)_{-3}$ & \\[.1in]
 \hline
 \end{tabular}\\[.2in]
 \caption{The Upper World. Chiral multiplets transform as shown
 under $SU_6\times SU_3\times SU_3'\times U_1$.
 Note that the $U_1$ factor here is a subgroup of the
 $SU_2$ factor which appears on the lower world.  This is a chiral spectrum
 which is completely free of all gauge and mixed anomalies.}
 \label{upper}
 \vspace{1.2in}
 \begin{tabular}{|c|c|}
 \hline
 &\\[-.1in]
 $(\,{\bf 10}\,,\,{\bf 3}\,,\,{\bf 1}\,)_{-2}$ &
 $16\,(\,{\bf 10}\,,\,{\bf 1}\,,\,{\bf 2}\,)_{0}$ \\[.1in]
 $(\,{\bf 1}\,,\,{\bf 3}\,,\,{\bf 1}\,)_{+4}$ &
 $16\,(\,{\bf 1}\,,\,{\bf 3}\,,\,{\bf 2}\,)_{+2}$ \\[.1in]
 $(\,{\bf 16}\,,\,{\bf 3}\,,\,{\bf 1}\,)_{+1}$ &
 $16\,(\,{\bf 1}\,,\,{\bf \bar{3}}\,,\,{\bf 2}\,)_{-2}$ \\[.1in]
 $(\,{\bf 16}\,,\,{\bf 1}\,,\,{\bf 1}\,)_{-3}$ &  \\[.1in]
 $(\,{\bf \bar{16}}\,,\,{\bf 1}\,,\,{\bf 1}\,)_{+3}$ & \\[.1in]
 \hline
 \end{tabular} \\[.2in]
 \caption{The Lower World. Chiral multiplets transform as shown
 under $SO_{10}\times SU_3'\times SU_2\times U_1'$.
 This is a chiral spectrum which is completely free of all
 gauge and mixed anomalies.}
 \label{lower}
 \end{center}
 \end{table}

 If we now take another limit, whereby the one remaining compact dimension
 $x^{11}$ shrinks to zero size, then the upper and lower worlds
 coalesce.  All fields then transform under
 $SO_{10}\times SU_6\times SU_3\times SU_3'\times U_1\times U_1'$,
 where $U_1\subset SU_2$.  The chiral spectrum is then obtained
 by combining (\ref{upper}) and (\ref{lower}) by considering the
 relevant branching rules.  For completeness, we include this
 spectrum in Table \ref{collided}.
 \begin{table}
 \begin{center}
 \begin{tabular}{|c|c|}
 \hline
 \hspace{1.5in} & \hspace{1.5in} \\[-.1in]
 $(\,{\bf 1}\,,\,{\bf 1}\,,\,{\bf 1}\,,\,{\bf 1}\,)_{-6,0}$ &
 $9\,(\,{\bf 1}\,,\,{\bf 6}\,,\,{\bf 1}\,,\,{\bf 3}\,)_{-1,0}$ \\[.1in]
 $(\,{\bf 1}\,,\,{\bf 15}\,,\,{\bf \bar{3}}\,,\,{\bf 1}\,)_{0,0}$ &
 $6\,(\,{\bf 1}\,,\,{\bf \bar{6}}\,,\,{\bf 1}\,,\,{\bf \bar{3}}\,)_{+1,0}$ \\[.1in]
 $(\,{\bf 1}\,,\,{\bf 20}\,,\,{\bf 1}\,,\,{\bf 1}\,)_{+3,0}$ &
 $9\,(\,{\bf 1}\,,\,{\bf 1}\,,\,{\bf 3}\,,\,{\bf 3}\,)_{+2,0}$ \\[.1in]
 $(\,{\bf 1}\,,\,{\bf 6}\,,\,{\bf 3}\,,\,{\bf 1}\,)_{-3,0}$ &
 $6\,(\,{\bf 1}\,,\,{\bf 1}\,,\,{\bf \bar{3}}\,,\,{\bf \bar{3}}\,)_{-2,0}$ \\[.1in]
 $(\,{\bf 1}\,,\,{\bf 6}\,,\,{\bf 3}\,,\,{\bf 1}\,)_{+3,0}$ & \\[.1in]
 $(\,{\bf 1}\,,\,{\bf \bar{6}}\,,\,{\bf \bar{3}}\,,\,{\bf 1}\,)_{-3,0}$ & \\[.1in]
 \hline
 &\\[-.1in]
 $(\,{\bf 10}\,,\,{\bf 1}\,,\,{\bf 1}\,,\,{\bf 3}\,)_{0,-2}$ &
 $16\,(\,{\bf 10}\,,\,{\bf 1}\,,\,{\bf 1}\,,\,{\bf 1}\,)_{+1,0}$ \\[.1in]
 $(\,{\bf 1}\,,\,{\bf 1}\,,\,{\bf 1}\,,\,{\bf 3}\,)_{0,+4}$ &
 $16\,(\,{\bf 10}\,,\,{\bf 1}\,,\,{\bf 1}\,,\,{\bf 1}\,)_{-1,0}$ \\[.1in]
 $(\,{\bf 16}\,,\,{\bf 1}\,,\,{\bf 1}\,,\,{\bf 3}\,)_{0,+1}$ &
 $16\,(\,{\bf 1}\,,\,{\bf 1}\,,\,{\bf 1}\,,\,{\bf 3}\,)_{+1,+2}$ \\[.1in]
 $(\,{\bf 16}\,,\,{\bf 1}\,,\,{\bf 1}\,,\,{\bf 1}\,)_{0,-3}$ &
 $16\,(\,{\bf 1}\,,\,{\bf 1}\,,\,{\bf 1}\,,\,{\bf 3}\,)_{-1,+2}$ \\[.1in]
 $(\,{\bf \bar{16}}\,,\,{\bf 1}\,,\,{\bf 1}\,,\,{\bf 1}\,)_{0,+3}$ &
 $16\,(\,{\bf 1}\,,\,{\bf 1}\,,\,{\bf 1}\,,\,{\bf \bar{3}}\,)_{+1,-2}$ \\[.1in]
 & $16\,(\,{\bf 1}\,,\,{\bf 1}\,,\,{\bf 1}\,,\,{\bf \bar{3}}\,)_{-1,-2}$ \\[.1in]
 \hline
 \end{tabular} \\[.2in]
 \caption{The four-dimensional spectrum seen when the upper and
 lower worlds coalesce, expressed in terms of $N=1$ chiral multiplets
 transforming under $SO_{10}\times SU_6\times SU_3\times SU_3'\times U_1\times U_1'$.
 Fields above the bar come from the upper world, while fields below the bar come
 from the lower world.  Those on the left are the survivors from ten-dimensional
 $E_8\times E_8$ fields.  Those on the right are the survivors of the six-dimensional
 twisted fields.}
 \label{collided}
 \end{center}
 \end{table}

 \section{Conclusions}
 We have made a microscopic analysis of the local anomaly
 cancellation requirements associated with a special {\it M}-theory orbifold.
 The construction we have studied is the simplest
 abelian quotient $T^7/\Gamma$
 which does not involve any freely acting
 involutions and which gives rise to a chiral
 four-dimensional spectrum.
 By demanding that all local anomalies at each point on each
 even-dimensional orbifold plane and orbifold-plane intersection
 vanish, we are able to determine a particular anomaly-free
 chiral spectrum associated with a pair of four-dimensional
 brane-worlds, linked by a five-dimensional bulk.

 A central part of our analysis relies on the group-theoretic
 restrictions related to what we have defined as
 ``consistently-entwined embeddings" of subgroups inside
 of subgroups.  We find it intriguing that these essentially
 crystallographic constraints emerge so naturally from
 intricate local anomaly considerations and, more-so, that
 these matters are so readily resolved.  We are engaged in applying
 these same techniques algorithmically to a systematic scan of
 a large class of {\it M}-theory orbifolds.  One purpose of this
 paper is to explain some of the core details of our algorithm, so that
 we can focus more exclusively on results in subsequent papers.
 We also find the details amusing.

 Owing to a comparative dearth of inroads, it remains
 relatively difficult to describe effective four-dimensional
 physics from a purely {\it M}-theoretic standpoint as compared to
 the situation in conventional string theory.
 For instance, in the case of string compactification schemes,
 detailed analyses of D-brane configurations on
 various orientifold backgrounds have allowed for a reasonably
 concise top-down approach towards the determination of chiral
 spectra, supersymmetry  breaking, and the computation of
 superpotentials.
 It remains somewhat mysterious how to
 determine all of the analogous data using what is yet
 known about {\it M}-theory.
 This fact is both a hindrance and a
 help.  It is helpful because it forces us to use the small
 amount of constraints available, mostly in the form of local
 anomaly conditions, for all they are worth.
 What is interesting, however, is just how snugly these conditions
 fit the problem.

 An open question is how to describe the {\it lift}
 of particular string compactification schemes to {\it M}-theory,
 if possible.  One simple known example is the case of
 the non-compactified $IIA$ string.  In this case,
 from the point of view of the effective theory,
 one may decompactify the $IIA$ supergravity theory
 by merely adding in a new circular dimension.  Another simple example
 is the case of the non-compactified $E_8\times E_8$
 heterotic string.  In that case
 one decompactifies the coupled $N=1$ supergravity-Yang-Mills
 theory by stretching a line segment out of each point in the originally
 ten-dimensional spacetime,
 keeping one $E_8$ sector
 on one ten-dimensional end-of-the-world and the other
 $E_8$ sector on the other ten-dimensional end-of-the-world.
 By way of contrast, the $SO_{32}$ string does not have such a
 direct {\it M}-theory lift.  This is related to the fact that
 the gauge group $SO_{32}$ cannot be consistently factorized;
 it resists being torn-apart.  In each of these cases, the subset
 of models which admit a direct lift corresponds to those which
 coincide with consistent {\it M}-theory compactifications.

 But what about more exotic situations?  Suppose one starts,
 for instance, with a $IIA$ string compactified on a particular orientifold,
 in the presence of a particular collection of D-branes
 and open strings.  A given scenario of this sort may or may not admit a
 clean lift to {\it M}-theory.  One method for probing
 this question is to compare the effective theory associated with
 a given string compactification scheme with the relevant set of consistent
 {\it M}-theory compactifications (assuming it is possible to delineate these).
 As regards the {\it M}-theory side of this issue,
 if we remain within the class of orbifold compactification schemes
 described in this paper,
 then the limitations on consistent effective descriptions
 correlate with the limited number of
 consistently-entwined embeddings of aggregate gauge lattices.
 Plausibly, these in turn correlate with subsets of the
 ways that one can consistently wrap D-branes on
 internal cycles of compactification spaces in string theory.
 In the {\it M}-theory approach one relies on group
 theory and crystallography, whereas in the D-brane picture one
 relies more heavily on the homology of the compactification spaces.
 It might be interesting to explore such relationships.

 One relevant observation is the appearance of various bi-fundamental
 representations in the effective {\it M}-theory spectra which we have
 derived.  From the D-brane point of view, these should arise from
 open strings stretched from
 one stack of D-branes to another.  We notice, however, in the {\it M}-theory
 model which we have derived, the appearance of
 other representations such as the higher antisymmetric
 tensors of $SU_n$  (the {\bf 15} and the {\bf 20}
 of $SU_6$, for instance).  It is less clear how to correlate these
 states with string theory analogues.
 It would be interesting to explore further the relationship
 between {\it M}-theoretic spectra and string-theoretic analogues.
 We expect that the models which we describe should
 descend to particular cases of $IIA$ string theory compactified
 on Calabi-Yau orientifolds with D-branes wrapping internal cycles
 of these spaces.  Among other things, we are actively probing
 such questions.

 \appendix

 \section{Four-Dimensional Anomaly Computation}
 \label{anom4d}
 In four dimensions, gauge anomalies appear in the presence of
 chiral spinor fields $\psi_R=\g_5\,\psi_R$.
 Assume the internal gauge group has $n$ simple factors and $m$ abelian
 $U(1)$ factors,
 \brr {\cal G}=\otimes_{I=1}^n\,{\cal G}_I\,\otimes_{l=1}^m\,U(1)_l \,.
 \err
 Assume, as well, that the chiral spinors can be described
 by $S$ sets of fields transforming according to
 \brr {\bf {\cal R}}=\oplus_{i=1}^S\,(\,{\bf{\cal R}}_{1\,i}\,,\,...\,,\,
      {\bf {\cal R}}_{n\,i}\,)_{q_{i\,1},...,q_{i\,m}}\,,
 \err
 where ${\bf {\cal R}}_{I\,i}$ describes the representation
 of the $i$th set of chiral fields in ${\cal G}_I$, and $q_{i\,l}$
 is the $l$th associated $U(1)$ charge.
 Anti-chiral spinors $\psi_L=-\g_5\,\psi_L$ transforming according to
 a representation ${\cal R}_i$ can be replaced by their charge conjugate spinors
 $\psi_R=C^{-1}\,\bar{\psi}^T_L$, which transform according to
 ${\bf \bar{\cal R}}_i$.  Without loss of generality, we therefore conventionally
 express fermion spectra
 exclusively in terms of chiral spinors, rather than as a mixture of chiral
 and anti-chiral spinors.
 In this case, the gauge anomaly is described, via descent equations,
 by the formal six-form $I({\rm GAUGE})_6={\rm tr}\,F^3$, where $F$ is the
 matrix-valued two-form field strength associated with ${\cal G}$.
 Gauge anomaly cancellation is equivalent to the requirement that the six-form
 $I({\rm GAUGE})_6$
 vanish.  This requires that each of the following
 numbers vanish,
 \brr I({\cal G}_{I})_l &\equiv&
      \sum_i\,\sum_{{\bf {\cal R}}_I}\,
      n(\,{\bf {\cal R}}_I\,)_i\,\,q_{i\,l}\,
      I_2(\,{\bf {\cal R}}\,)
      \nonumber\\[.1in]
      I(\,U_1\,)_{l\,,\,{\rm GAUGE}} &=&
      \sum_i\,N_i\,q_{i\,l}^3 \,,
 \label{ganom}\err
 where $I_2(\,{\bf {\cal R}}_I\,)$ is the second index of the
 representation ${\bf {\cal R}}_I$ associated with the $I$th simple
 factor ${\cal G}_I$, $n(\,{\bf {\cal R}}_I\,)_i$ is the
 multiplicity of fields in ${\bf {\cal R}}_i$ transforming in the representation
 ${\bf {\cal R}}_I$ of ${\bf {\cal G}}_I$, and $N_i$ are the total
 number of fields in ${\bf {\cal R}}_i$.

 We are also interested in the gauge/gravitational ``mixed" anomaly.
 This anomaly is related by descent, to the
 formal six form $I({\rm MIXED})_6={\rm tr}\,R^2\wedge {\rm tr}\,F$.
 Mixed anomaly cancellation is equivalent to requiring that the
 six-form $({\rm MIXED})_6$ vanish.  This is equivalent to
 requiring that the following numbers vanish,
 \brr I(\,U_1\,)_{l\,{\rm MIXED}} &=&
      \sum_i\,N_i\,q_{i\,l} \,.
 \label{manom}\err
 Thus, mixed anomaly cancellation requires that, for each $U(1)$
 factor, the sum of all the charges vanish.
 (Note that, taken together, gauge and mixed anomaly cancellation require
 the sum of the charges and also the sum of the {\it cubes} of the
 charges vanish.)

 \section{Indices for $SU(N)$}
 \label{indices}
 Each representation of a classical Lie algebra has a set of
 associated rational indices.  For instance, the second
 index of a representation ${\bf {\cal R}}$ is defined by the
 relationship
 \brr {\rm tr}_{\bf {\cal R}}\,F^2=I_2({\bf\cal R})\,{\rm tr}\,F^2 \,,
 \err
 where the trace on the left-hand side is over the representation
 ${\bf {\cal R}}$ and the trace on the right-hand side is
 over the fundamental representation.
 There is a useful and concise algorithm, derived in
 \cite{rsv}, for determining representation indices
 for any antisymmetric tensor representation of $SU(N)$.
 For instance, the 2nd index $I_2([k])$ for all of the $[k]$ representations
 of $SU(N)$ are encapsulated in the polynomial
 \brr P_n^N(x)=-(1+x)^N\,\sum_{l=1}^\infty\,l\,(-x)^l
 \label{poly}\err
 and are read off by the indentification
 \brr P_2^N(x)=\sum_{k=1}^\infty I_n([k])\,x^k \,.
 \err
 So, in order to determine the index $I_2([k])$ for the group
 $SU(N)$, for given $k$ and $N$, we first compute the polynomial
 $P_n^N(x)$ using (\ref{poly}), and then read off the coefficient
 of $x^k$.  That number is $I_2([k])$. Note that the second index
 of the fundamental $[1]$ representation is always unity,
 $I_2([1])=1$.

 \vspace{.5in}
{\Large {\bf Acknowledgements}}\\[.1in]
 M.F. would like to thank Dieter L{\" u}st for helpful comments and for
 warm hospitality at Humboldt University, where a portion of this manuscript was
 prepared, and also Burt Ovrut for discussions.

 \end{document}